\documentclass[english,eqsecnum,nofootinbib,superscriptaddress]{revtex4-2}
\usepackage[T1]{fontenc}
\usepackage[latin9]{inputenc}
\usepackage{color}
\usepackage{babel}
\usepackage{mathtools}
\usepackage{bm}
\usepackage{amsmath}
\usepackage{geometry}
\usepackage[pdfusetitle,
 bookmarks=true,bookmarksnumbered=false,bookmarksopen=false,
 breaklinks=false,pdfborder={0 0 1},backref=false,colorlinks=true]
 {hyperref}

\makeatletter
\usepackage{datetime2}
\usepackage{extarrows}

\makeatother

\begin{document}
\title{Spatially covariant gravity with two degrees of freedom in the presence
of an auxiliary scalar field: Hamiltonian analysis}
\author{Jun-Cheng Zhu}
\affiliation{School of Physics and Astronomy, Sun Yat-sen University, Zhuhai 519082,
China}
\author{Shu-Yu Li}
\affiliation{School of Physics and Astronomy, Sun Yat-sen University, Zhuhai 519082,
China}
\author{Xian Gao}
\email[Corresponding author: ]{gaoxian@mail.sysu.edu.cn}

\affiliation{School of Physics, Sun Yat-sen University, Guangzhou 510275, China}
\begin{abstract}
A class of gravity theories respecting spatial covariance and in the
presence of non-dynamical auxiliary scalar fields with only spatial
derivatives is investigated. Generally, without higher temporal derivatives
in the metric sector, there are 3 degrees of freedom (DOFs) propagating
due to the breaking of general covariance. Through a Hamiltonian constraint
analysis, we examine the conditions to eliminate the scalar DOF such
that only 2 DOFs, which correspond the tensorial gravitational waves
in a homogeneous and isotropic background, are propagating. We find
that two conditions are needed, each of which can eliminate half degree
of freedom. The second condition can be further classified into two
cases according to its effect on the Dirac matrix. We also apply the
formal conditions to a polynomial-type Lagrangian as a concrete example,
in which all the monomials are spatially covariant scalars containing
two derivatives. Our results are consistent with the previous analysis
based on the perturbative method.
\end{abstract}
\maketitle

\section{Introduction}

In light of the detection of gravitational waves, gravitational waves
provide a novel tool for testing theories of gravity. General relativity
possesses only two propagating degrees of freedom (DOFs), which manifests
as tensorial gravitational waves (GWs) in a homogeneous and isotropic
background. The detection of GWs raises a fundamental question: Is
general relativity the only theory of gravity that propagates GWs
with precisely two degrees of freedom (2DOFs)? This question is partially
answered by Lovelock's theorem \citep{Lovelock:1971yv,Lovelock:1972vz},
which states that in four dimensions, requiring spacetime covariance
and second-order equations of motion, general relativity is indeed
the unique theory propagating gravitational waves with two degrees
of freedom. This also implies that, by violating the assumptions of
Lovelock's theorem, other theories of gravity propagating 2DOFs can,
in principle, exist. From the theoretical point of view, a Lorentz
covariant massless spin-2 particle possesses two physical polarization
states. If one constructs a Lorentz invariant theory for such a massless
spin-2 particle and further demands that it can be consistently coupled
to matter fields, the resulting structure is necessarily that of general
relativity. In this sense, general relativity as a theory of gravity
is unique since its form is inextricably related to the masslessness
of the graviton and the Lorentz symmetry of spacetime. Consequently,
alternative theories of gravity that also propagate only 2DOFs provide
a valuable testing ground for the foundation principles of general
relativity, such as Lorentz invariance, diffeomorphism invariance,
and the masslessness of graviton, etc.

Gravitational theories propagating 2DOFs can be traced back to the
Cuscuton theory \citep{Afshordi:2006ad}, which is introduced as a
scalar field model with a non-canonical kinetic term $\propto\sqrt{\left|-\partial_{\mu}\phi\partial^{\mu}\phi\right|}$.
The cuscuton theory represents the incompressible (infinite speed
of sound) limit of $k$-essence, in which the scalar perturbation
propagates with an infinite speed of sound in a cosmological background.
Although linear perturbations exhibit superluminal propagation, the
local phase space degenerates, implying no independent dynamical DOFs
and thus preserving causality. As a result, the cuscuton theory modifies
gravitational dynamics without introducing additional local DOFs.
The Cuscuton theory was extensively studied \citep{Afshordi:2007yx,Mylova:2023ddj,Boruah:2017tvg,Boruah:2018pvq,Bhattacharyya:2016mah,Quintin:2019orx,Bartolo:2021wpt,HosseiniMansoori:2022xnq,Channuie:2023ddv,Maeda:2022ozc,Kohri:2022vst,Panpanich:2021lsd,Lima:2023str}
and was extended in the framework of scalar-tensor theory with higher
order derivatives \citep{Iyonaga:2018vnu,Iyonaga:2020bmm}. Such kind
of theory has also been discussed as a special case of Ho\v{r}ava
gravity \citep{Afshordi:2009tt,Zhu:2011xe,Zhu:2011yu,Chagoya:2018yna}. The relation between the 2DOFs and the spacetime
symmetry has been investigated in \citep{Khoury:2011ay,Khoury:2014sea,Chagoya:2016inc,Tasinato:2020fni}. Another class
of of theories propagating 2DOFs was proposed in \citep{Lin:2017oow},
which is dubbed the minimally modified gravity (MMG)\footnote{See also the minimal theory of massive gravity studied in \citep{DeFelice:2015hla,DeFelice:2015moy,Bolis:2018vzs,DeFelice:2018vza,DeFelice:2021trp,DeFelice:2023bwq}.}. As the name suggests, MMG theories modify generally relativity ``minimally''
in the sense that they modify the gravitational sector without introducing
extra DOFs beyond the standard 2DOFs. MMG theories extend general
relativity by breaking time diffeomorphism invariance while preserving
spatial diffeomorphism invariance, and whose action is linear in the
lapse function. A key result is the derivation of a self-consistency
condition, which ensures the theory possesses no more than two local
physical DOFs. Examples of MMG include general relativity, square-root
gravity, exponential gravity, and theories with lapse-independent
terms, all of which exhibit a constrained phase space structure that
eliminates the scalar graviton through first- or second-class constraints.
The so-called type-II MMG \citep{Aoki:2018brq,Aoki:2018zcv}
includes the original Cuscuton theory as a special case. The MMG has
been studied extensively \citep{Mukohyama:2019unx,DeFelice:2020eju,DeFelice:2020cpt,Aoki:2020oqc, Pookkillath:2021gdp,DeFelice:2020ecp,Aoki:2021zuy,DeFelice:2020onz,DeFelice:2020prd,DeFelice:2021xps,DeFelice:2022uxv,Jalali:2023wqh,Carballo-Rubio:2018czn,Sangtawee:2021mhz,Ganz:2022iiv,Akarsu:2024qsi}. It has also been generalized
in \citep{Yao:2020tur,Yao:2023qjd} by introducing auxiliary
constraints in the phase space.

A general framework of spatially covariant gravity (SCG) theories
respecting only the spatial covariance was proposed in \citep{Gao:2014soa,Gao:2014fra}. Due to the violation of general covariance,
SCG generally propagates a scalar degree of freedom (besides the 2DOFs
corresponding to the gravitational waves) and thus can be viewed as
an alternative approach to constructing the scalar-tensor theories
\citep{Gao:2020juc,Gao:2020yzr,Gao:2020qxy,Hu:2021bbo}. The SCG includes the Horava gravity \citep{Horava:2008ih,Horava:2009uw,Blas:2009yd}, effective field theory
of inflation/dark energy \citep{Creminelli:2006xe,Cheung:2007st,Creminelli:2008wc,Gubitosi:2012hu,Bloomfield:2012ff}
as special cases. It was further generalized with a dynamical lapse
function \citep{Gao:2018znj,Gao:2019lpz}, with nonmetricity
\citep{Yu:2024drx}, with parity violation \citep{Hu:2024hzo} and
with multiple scalar modes \citep{Yu:2024sed}. The SCG has been applied
in the study of cosmology and gravitational waves \citep{Fujita:2015ymn,Gao:2019liu,Zhu:2022dfq,Zhu:2022uoq,Zhu:2023rrx}. Subclasses of SCG propagating only 2DOFs were
explored in \citep{Gao:2019twq,Hu:2021yaq} and with a dynamical
lapse function in \citep{Lin:2020nro}. The concrete Lagrangian found
in \citep{Gao:2019twq} has been applied in the study of cosmology
\citep{Iyonaga:2021yfv,Hiramatsu:2022ahs,Saito:2023bhn,Bartolo:2021wpt,Chakraborty:2023jek}.

In this work, we will generalize the SCG in the presence of an auxiliary
scalar field, and determine the conditions such that the theory propagates
only 2DOFs. The idea of spatially covariant gravity with an auxiliary
scalar field was firstly introduced in \citep{Gao:2018izs}, which
was originally motivated by the generally covariant scalar-tensor
theory when the scalar field possess a spacelike gradient\footnote{However, we would like to point out that they are completely different
theories. Please refer to Appendix \ref{app:comp} for details. }. The presence of an auxiliary scalar field gives us more and novel
possibilities to build the theory propagating two degrees of freedom. 

In \citep{Wang:2024hfd}, by performing a perturbative analysis similar
to \citep{Iyonaga:2018vnu,Gao:2019lpz,Hu:2021yaq},
necessary conditions such that no scalar mode propagates at linear
order in perturbations around a cosmological background were determined.
Nevertheless a Hamiltonian constraint analysis is still needed in
order to determine the necessary and sufficient conditions for the
theory to propagate only 2DOFs at the nonlinear level. This paper
is thus devoted to this issue.

This paper is organized as follows. In Sec. \ref{sec:scg}, we describe
our model of spatially covariant gravity with an auxiliary scalar
field. It was found that new terms in the Lagrangian are allowed thanks
to the introduction of the auxiliary scalar field. In Sec. \ref{sec:ham},
we derive the Hamiltonian formalism and show that the theory generally
propagates a scalar mode if no further constraint is imposed. In Sec.
\ref{sec:deg}, we make the degeneracy analysis and derive the conditions
such that the theory propagate only 2DOFs. In Sec. \ref{sec:mod},
we apply our formal analysis to a concrete model, of which the Lagrangian
is built of SCG monomials of $d=2$. In Sec. \ref{sec:con}, we summarize
our results.

\section{Spatially covariant gravity with an auxiliary scalar field \label{sec:scg}}

Our starting point is the action
\begin{equation}
S=\int\mathrm{d}t\mathrm{d}^{3}xN\sqrt{h}\mathcal{L}\left(\phi,N,h_{ij},{}^{3}\!R_{ij};\mathrm{D}_{i},\pounds_{\bm{n}}\right),\label{action}
\end{equation}
where $N$ is the usual lapse function, $h_{ij}$ is the spatial metric
on the spacelike hypersurfaces, $^{3}\!R_{ij}$ is the spatial Ricci
tensor, $\mathrm{D}_{i}$ is the spatially covariant derivative compatible
with $h_{ij}$, and $\pounds_{\bm{n}}$ is the Lie derivative with
respect to the normal vector $n^{a}$ of the hypersurfaces. By definition,
the action (\ref{action}) respects only the 3-dimensional spatial
invariance. We require that the Lie derivative $\pounds_{\bm{n}}$
acts only on $h_{ij}$, which yields the extrinsic curvature $K_{ij}$
defined by
\begin{equation}
\pounds_{\boldsymbol{n}}h_{ij}=2K_{ij}=\frac{1}{N}\left(\dot{h}_{ij}-\pounds_{\vec{N}}h_{ij}\right).\label{extK}
\end{equation}
As a result, both the lapse function $N$ and the scalar field $\phi$
play the role of an auxiliary variables. Note the shift vector $N^{i}$
should not appear explicitly in the Lagrangian, which is actually
the gauge field of the spatial covariance.

In this work, we restrict our attention to the first-order Lie derivative
to prevent higher-order time derivatives from appearing in the equations
of motion. Moreover, since the Lie derivative is assumed to act solely
on $h_{ij}$, it enters the Lagrangian exclusively through the extrinsic
curvature (\ref{extK}). Therefore, we simply introduce $K_{ij}$
to replace the Lie derivative $\pounds_{\bm{n}}$ in (\ref{action}),
and consider the following action
\begin{equation}
\tilde{S}=\int\mathrm{d}t\mathrm{d}^{3}x\,N\sqrt{h}\mathcal{L}\left(\phi,N,h_{ij},{}^{3}\!R_{ij},K_{ij};\mathrm{D}_{i}\right).\label{actionK}
\end{equation}
According to the Hamiltonian analysis in the next section, generally
the action (\ref{actionK}) has 3 DOFs. This is simply because $\phi$
has no dynamics, the number of DOFs of (\ref{actionK}) is the same
as the spatially covariant gravity without the scalar field \citep{Gao:2014soa,Gao:2014fra}.
In particular, since the extrinsic curvature is linear in the Lie
derivative (and thus in time derivative), the resulting equations
of motion from the action (\ref{actionK}) contain at most second-order
time derivatives. As a result, the theory avoids the Ostrogradsky
ghost instability.

In principle, Lie derivatives of the lapse function $\pounds_{\bm{n}}N$
and the extrinsic curvature $\pounds_{\bm{n}}K_{ij}$ can be considered,
which will introduced more degrees of freedom. It was shown that extra
conditions must be put in order to evade these unwanted degrees of
freedom. For example, the theory will propagate an extra DOF (i.e.,
in total 4 DOFs) in the case of a dynamical lapse function. Two conditions
are needed in order to evade such an extra DOF \citep{Gao:2018znj}.
Similar analysis was considered in order to fully eliminate the scalar
DOFs \citep{Lin:2020nro}. 

Generally, the Lagrangian in (\ref{actionK}) may be nonlinear in
$K_{ij}$, which makes the explicit reversion of velocity $\dot{h}_{ij}$
in terms of the momentum $\pi^{ij}$ impossible. This problem can
be solved by rewriting (\ref{actionK}) in an equivalent form
\begin{align}
\tilde{S} & =\int\mathrm{d}t\mathrm{d}^{3}xN\sqrt{h}\tilde{\mathcal{L}}\left(\phi,N,h_{ij},{}^{3}\!R_{ij},K_{ij};\mathrm{D}_{i}\right)\nonumber \\
 & =S\left(\phi,N,h_{ij},{}^{3}\!R_{ij},B_{ij};\mathrm{D}_{i}\right)+\int\mathrm{d}t\mathrm{d}^{3}xN\sqrt{h}\varLambda^{ij}\left(K_{ij}-B_{ij}\right),\label{actionLM}
\end{align}
where the action $S\left(\phi,N,h_{ij},{}^{3}R_{ij},B_{ij};\mathrm{D}_{i}\right)$
is the same action as (\ref{actionK}) with $K_{ij}$ replaced by
$B_{ij}$. This strategy of using the Lagrange multiplier to linearize
the velocities has been used in \citep{Gao:2018znj,Gao:2019twq}.
The equations of motion for $\varLambda^{ij}$ ensure the auxiliary
field $B_{ij}$ to be exactly identical to $K_{ij}$. Thus the two
actions are equivalent, at least at the classical level. By making
use of the equation of motion for the auxiliary field $B_{ij}$, we
can fix the Lagrangian multiplier $\varLambda^{ij}$ as
\begin{equation}
\varLambda^{ij}=\frac{1}{N\sqrt{h}}\frac{\delta S}{\delta B_{ij}}.\label{solLM}
\end{equation}

Finally, by plugging the solution of $\varLambda^{ij}$, the action
(\ref{actionLM}) can be recast in the form
\begin{equation}
\tilde{S}=S\left(\phi,N,h_{ij},{}^{3}\!R_{ij},B_{ij};\mathrm{D}_{i}\right)+\int\mathrm{d}t\mathrm{d}^{3}x\frac{\delta S}{\delta B_{ij}}\left(K_{ij}-B_{ij}\right).\label{action_fin}
\end{equation}
In the following, we will use (\ref{action_fin}) as our starting
point for the Hamiltonian analysis.

\section{Hamiltonian formalism \label{sec:ham}}

\subsection{Hamiltonian and primary constraints}

In the action (\ref{action_fin}), the set of variables is
\begin{equation}
\varPhi^{I}\coloneqq\left\{ N^{i},\phi,B_{ij},N,h_{ij}\right\} ,\label{var}
\end{equation}
and the set of their conjugate momenta is denoted by
\begin{equation}
\varPi_{I}\coloneqq\left\{ \pi_{i},p,p^{ij},\pi,\pi^{ij}\right\} .\label{mom}
\end{equation}

The conjugate momenta are defined as
\begin{align}
\varPi_{I} & \coloneqq\frac{\delta\tilde{S}}{\delta\dot{\Phi}^{I}}.\label{momdef}
\end{align}
Since all the velocities $\dot{\varPhi}^{I}$ in the Lagrangian (\ref{action_fin})
are linearized, the conjugate momenta autonatically correspond to
constraints among phase space variables. Indeed, the conjugate momenta
are
\begin{eqnarray}
\left\{ \pi_{i},p,p^{ij},\pi,\pi^{ij}\right\}  & = & \left\{ \frac{\delta\tilde{S}}{\delta\dot{N}^{i}},\frac{\delta\tilde{S}}{\delta\dot{\phi}},\frac{\delta\tilde{S}}{\delta\dot{B}_{ij}},\frac{\delta\tilde{S}}{\delta\dot{N}},\frac{\delta\tilde{S}}{\delta\dot{h}_{ij}}\right\} \nonumber \\
 & = & \left\{ 0^{i},0,0_{ij},0,\frac{1}{2N}\frac{\delta S}{\delta B_{ij}}\right\} ,\label{momxpl}
\end{eqnarray}
which result in the primary constraints
\begin{equation}
\pi_{i}\approx0,\quad p\approx0,\quad p^{ij}\approx0,\quad\pi\approx0,\quad\tilde{\pi}^{ij}\equiv\pi^{ij}-\frac{1}{2N}\frac{\delta S}{\delta B_{ij}}\approx0.\label{pricons}
\end{equation}
Here and throughout this work \textquotedblleft $\approx$\textquotedblright{}
represents \textquotedblleft weak equality\textquotedblright , i.e.,
an equality that holds only when the constraints are satisfied. On
the other hand, the strong equality \textquotedblleft $=$\textquotedblright{}
holds no matter the constraints are satisfied or not. 

We denote the set of primary constraints as
\begin{equation}
\varphi^{I}\coloneqq\left\{ \pi_{i},p,p^{ij},\pi,\tilde{\pi}^{ij}\right\} ,\label{priconsphi}
\end{equation}
and the subspace of the phase space when the primary constraints are
satisfied as $\varGamma_{\mathrm{P}}$, which can be viewed as a hypersurface
in the original phase space $\left\{ \varPhi^{I},\varPi_{I}\right\} $.

The canonical Hamiltonian is obtained by performing the Legendre transformation
\begin{align}
H_{\mathrm{C}} & =\int\mathrm{d}^{3}x\left(\sum_{I}\varPi_{I}\dot{\varPhi}^{I}-N\sqrt{h}\tilde{\mathcal{L}}\right)\approx\left.H_{\mathrm{C}}\right|_{\varGamma_{\mathrm{P}}},\label{HamC}
\end{align}
where
\begin{equation}
\left.H_{\mathrm{C}}\right|_{\varGamma_{\mathrm{P}}}=\int\mathrm{d}^{3}x(NC)+X[\vec{N}],\label{HamC_GP}
\end{equation}
with
\begin{equation}
C=2\pi^{ij}B_{ij}-\sqrt{h}\mathcal{L}.\label{Cdef}
\end{equation}
For a general ``spatial'' vector field $\vec{\xi}$, $X[\vec{\xi}]$
is defined as \citep{Gao:2018znj}
\begin{eqnarray}
X[\vec{\xi}] & \coloneqq & \int\mathrm{d}^{3}x\,\varPi_{I}\pounds_{\vec{\xi}}\varPhi^{I}\nonumber \\
 & = & \int\mathrm{d}^{3}x\left(\pi_{i}\pounds_{\vec{\xi}}N^{i}+p\pounds_{\vec{\xi}}\phi+p^{ij}\pounds_{\vec{\xi}}B_{ij}+\pi\pounds_{\vec{\xi}}N+\pi^{ij}\pounds_{\vec{\xi}}h_{ij}\right),\label{Xdef}
\end{eqnarray}

In the following we briefly mention the properties of $X[\vec{\xi}]$
following \citep{Gao:2018znj}. Supposing $Q\left(\vec{x}\right)$
is a general spatial tensor field or density (with spatial indices
suppressed for briefness) built of phase space variables $\left\{ \varPhi^{I},\varPi_{I}\right\} $,
and $\mathcal{F}$ is a scalar functional defined by
\begin{equation}
\mathcal{F}\left[f,\varPhi^{I},\varPi_{I}\right]=\int\mathrm{d}^{3}x\,f(\vec{x})Q(\vec{x}),\label{testfun}
\end{equation}
with $f\left(\vec{x}\right)$ is a smoothing function. Note $f(\vec{x})$
also constains appropriate spatial indices (suppressed as well) such
that $\mathcal{F}$ is invariant under time-independent spatial diffeomorphism.
The Poisson bracket of $X[\vec{\xi}]$ and $\mathcal{F}$ are
\begin{eqnarray}
[X[\vec{\xi}],\mathcal{F}] & = & X\left[[\vec{\xi},\mathcal{F}]\right]+\int\mathrm{d}^{3}x\frac{\delta\mathcal{F}}{\delta f}\pounds_{\vec{\xi}}f\nonumber \\
 & = & \int\mathrm{d}^{3}x\,Q(\vec{x})\pounds_{\vec{\xi}}f(\vec{x}).\label{PB_XcalF}
\end{eqnarray}
In (\ref{PB_XcalF}), the Poisson bracket $[\mathcal{F},\mathcal{G}]$
is defined by
\begin{equation}
[\mathcal{F},\mathcal{G}]\coloneqq\int\mathrm{d}^{3}z\left(\frac{\delta\mathcal{F}}{\delta\varPhi_{I}(\vec{z})}\frac{\delta\mathcal{G}}{\delta\varPi^{I}(\vec{z})}-\frac{\delta\mathcal{F}}{\delta\varPi^{I}(\vec{z})}\frac{\delta\mathcal{G}}{\delta\varPhi_{I}(\vec{z})}\right).\label{PBdef}
\end{equation}
The key point is that the right side of the second line of the equation
vanishes on the constraint surface if $Q(\vec{x})$ is a constraint.

Through integrations by parts, $X[\vec{\xi}]$ can be written in a
more convenient form
\begin{equation}
X[\vec{\xi}]=\int\mathrm{d}^{3}x\,\xi^{i}\mathcal{C}_{i},\label{X_convnt}
\end{equation}
where
\begin{eqnarray}
\mathcal{C}_{i} & = & \pi\mathrm{D}_{i}N-2\sqrt{h}\mathrm{D}_{j}\left(\frac{1}{\sqrt{h}}\pi_{i}^{j}\right)+p\mathrm{D}_{i}\phi+p^{kl}\mathrm{D}_{i}B_{kl}\nonumber \\
 &  & -2\sqrt{h}\mathrm{D}_{j}\left(\frac{p^{jk}}{\sqrt{h}}B_{ik}\right)+\pi_{j}\mathrm{D}_{i}N^{j}+\sqrt{h}\mathrm{D}_{j}\left(\frac{1}{\sqrt{h}}\pi_{i}N^{j}\right).\label{calCi}
\end{eqnarray}
As we shall see later, $\mathcal{C}_{i}\approx0$ are actually the
secondary constraints associated with the primary constraints $\pi_{i}\approx0$.
Moreover, the Poisson bracket of $\mathcal{C}_{i}$ with any constraint
$Q$ can be obtained by
\begin{equation}
\left[X[\vec{\xi}],\mathcal{F}\right]=\int\mathrm{d}^{3}x\mathrm{d}^{3}y\,\xi^{i}(x)f(y)\left[\mathcal{C}_{i}(\vec{x}),Q(\vec{y})\right].\label{PBX}
\end{equation}
It immediately follows from (\ref{PB_XcalF}) that
\begin{equation}
\left[\mathcal{C}_{i}(\vec{x}),Q(\vec{y})\right]\approx0,\quad\text{for any}\quad Q\approx0.
\end{equation}

Due to the presence of primary constraints, the time evolution is
determined by the total Hamiltonian defined by
\begin{equation}
H_{\mathrm{T}}\coloneqq H_{\mathrm{C}}+\int\mathrm{d}^{3}y\sum_{I}\lambda_{I}(\vec{y})\varphi^{I}(\vec{y}),\label{totHam}
\end{equation}
where $\lambda_{I}:=\left\{ v^{i},v,v_{ij},\lambda,\lambda_{ij}\right\} $
are the undetermined Lagrange multipliers. Since the canonical Hamiltonian
$H_{\mathrm{C}}$ is well-defined only on the subspace defined by
the primary constraints $\varGamma_{\mathrm{P}}$, we can directly
use $\left.H_{\mathrm{C}}\right|_{\varGamma_{\mathrm{P}}}$ given
in (\ref{HamC_GP}) instead of $H_{\mathrm{C}}$ in (\ref{HamC})
in the subsequent calculations.

\subsection{Consistency condition}

Constraints must be preserved in time evolution. For the primary constraints,
we must require that
\begin{equation}
\int\mathrm{d}^{3}y\left[\varphi^{I}(\vec{x}),\varphi^{J}(\vec{y})\right]\lambda_{J}(\vec{y})+\left[\varphi^{I}(\vec{x}),H_{\mathrm{C}}\right]\approx0,\label{conscond}
\end{equation}
which are the so-called consistency conditions for the primary constraints.
These conditions may yields further constraints. 

With no further requirement on the structure of the Lagrangian in
(\ref{action_fin}), the Poisson brackets between primary constraints
have been calculated in detail in \citep{Gao:2018izs}, which are
summarized in Appendix \ref{app:cons}. For later convenience, we
write (\ref{conscond}) in matrix form
\begin{equation}
\int\mathrm{d}^{3}y\left(\begin{array}{ccccc}
0 & 0 & 0 & 0 & 0\\
0 & 0 & 0 & 0 & \left[p(\vec{x}),\tilde{\pi}^{kl}(\vec{y})\right]\\
0 & 0 & 0 & 0 & \left[p^{ij}(\vec{x}),\tilde{\pi}^{kl}(\vec{y})\right]\\
0 & 0 & 0 & 0 & \left[\pi(\vec{x}),\tilde{\pi}^{kl}(\vec{y})\right]\\
0 & \left[\tilde{\pi}^{ij}(\vec{x}),p(\vec{y})\right] & \left[\tilde{\pi}^{ij}(\vec{x}),p^{kl}(\vec{y})\right] & \left[\tilde{\pi}^{ij}(\vec{x}),\pi(\vec{y})\right] & \left[\tilde{\pi}^{ij}(\vec{x}),\tilde{\pi}^{kl}(\vec{y})\right]
\end{array}\right)\left(\begin{array}{c}
v^{i}\\
v\\
v_{ij}\\
\lambda\\
\lambda_{ij}
\end{array}\right)\approx\left(\begin{array}{c}
-\mathcal{C}_{i}(\vec{x})\\
\frac{\delta S}{\delta\phi(\vec{x})}\\
0\\
-\mathcal{C}(\vec{x})\\
\left[\tilde{\pi}^{ij}(\vec{x}),H_{\mathrm{C}}\right]
\end{array}\right),\label{conscond_mat}
\end{equation}
where $\mathcal{C}_{i}$ is defined in (\ref{PBX}). Note $\mathcal{C}_{i}$
must have vanishing Poisson brackets with any other constraints due
to the property (\ref{PBX}). In the fourth line of (\ref{conscond_mat}),
$\mathcal{C}$ is defined by
\begin{equation}
\mathcal{C}=2\pi^{ij}(\vec{x})B_{ij}(\vec{x})-\frac{\delta S}{\delta N(\vec{x})},\label{calC}
\end{equation}
and $\left[\tilde{\pi}^{ij}(\vec{x}),H_{\mathrm{C}}\right]$ in the
fifth line is given by
\begin{equation}
\left[\tilde{\pi}^{ij}(\vec{x}),H_{\mathrm{C}}\right]=\frac{\delta S}{\delta h_{ij}(\vec{x})}-\frac{1}{N(\vec{x})}\int\mathrm{d}^{3}z\frac{\delta^{2}S}{\delta B_{ij}(\vec{x})\delta h_{kl}(\vec{z})}N(\vec{z})B_{kl}(\vec{z}).\label{PB_pitijHC_def}
\end{equation}

According to (\ref{PB_pijpitkl}), $\left[p^{ij}(\vec{x}),\tilde{\pi}^{kl}(\vec{y})\right]$
in the third line in (\ref{conscond_mat}) is proportional to the
kinetic matrix
\begin{equation}
\frac{\delta^{2}S_{B}}{\delta B_{ij}(\vec{x})\delta B_{kl}(\vec{y})},\label{kinmat}
\end{equation}
which we assume to be non-degenerate. This is because $B_{ij}$ is
the auxiliary field corresponding to $K_{ij}$, The non-degeneracy
of (\ref{kinmat}) implies that the kinetic term for $\dot{h}_{ij}$
is not degenerate, which guarantees the theory includes General Relativity
as its limit, and in particular, the existence of gravitational waves. 

As a result, the Lagrange multiplies $\lambda_{ij}$ can be fixed
from the third line in (\ref{conscond_mat}) to be
\begin{equation}
\lambda_{ij}=0.\label{lmbdij0}
\end{equation}
Then the second and the fourth line of (\ref{conscond_mat}) yield
another two secondary constraints
\begin{equation}
\frac{\delta S}{\delta\phi(\vec{x})}\approx0,\quad\mathcal{C}(\vec{x})\approx0.\label{seccons}
\end{equation}
Constraints $\frac{\delta S}{\delta\phi(\vec{x})}$ and $\mathcal{C}(\vec{x})$
do not generate additional constraints due to consistency conditions,
which will be clear in the next section.

After finding all the primary and secondary constraints, we are now
ready to classify all the constraints. According to the terminology
of Dirac-Bergmann algorithm, a constraint that has vanishing Poisson
brackets with all contraints (including itself) is dubbed a first-class
constraint, otherwise a second-class constraint. The difference is
that each first-class constraint eliminates a single degree of freedom,
while a second-class constraint, which reduce one phase space dimension,
eliminate half degree of freedom.

According to the above discussion on $\mathcal{C}_{i}$ and constraint
algebra in Appendix \ref{app:cons}, the Poisson brackets among all
the constraints can be summarized in the so-called Dirac matrix:
\begin{equation}
\begin{array}{|c|cccccccc|}
\hline [\cdot,\cdot] & \pi_{i}(\vec{y}) & \mathcal{C}_{i}(\vec{y}) & p(\vec{y}) & \pi(\vec{y}) & p^{kl}(\vec{y}) & \tilde{\pi}^{kl}(\vec{y}) & \mathcal{C}(\vec{y}) & \frac{\delta S}{\delta\phi(\vec{y})}\\
\hline \pi_{i}(\vec{x}) & 0 & 0 & 0 & 0 & 0 & 0 & 0 & 0\\
\mathcal{C}_{i}(\vec{x}) & 0 & 0 & 0 & 0 & 0 & 0 & 0 & 0\\
p(\vec{x}) & 0 & 0 & 0 & 0 & 0 & X & X & X\\
\pi(\vec{x}) & 0 & 0 & 0 & 0 & 0 & X & X & X\\
p^{ij}(\vec{x}) & 0 & 0 & 0 & 0 & 0 & X & X & X\\
\tilde{\pi}^{ij}(\vec{x}) & 0 & 0 & X & X & X & X & X & X\\
\mathcal{C}(\vec{x}) & 0 & 0 & X & X & X & X & X & X\\
\frac{\delta S}{\delta\phi(\vec{x})} & 0 & 0 & X & X & X & X & X & 0
\\\hline \end{array},
\end{equation}
where a \textquotedblleft $X$\textquotedblright{} stands for generally
non-vanishing Poisson brackets. In total there are 22 constraints
which can be divided into two classes:
\[
\begin{array}{rl}
6\text{ first-class: } & \pi_{i},\quad\mathcal{C}_{i},\\
16\text{ second-class: } & p,\quad p^{ij},\quad\pi,\quad\tilde{\pi}^{ij},\quad\frac{\delta S}{\delta\phi},\quad\mathcal{C}.
\end{array}
\]
The number of DOF is thus
\begin{eqnarray}
\#_{\mathrm{DOF}} & = & \frac{1}{2}\left(2\times\#_{\mathrm{var}}-2\times\#_{1\mathrm{st}}-\#_{2\mathrm{nd}}\right)\nonumber \\
 & = & \frac{1}{2}(2\times17-2\times6-16)\nonumber \\
 & = & 3,\label{NumDOF3}
\end{eqnarray}
which explicit shows that spatially covariant gravity with temporal
derivative arises only through the extrinsic curvature, has 3 propagating
degrees of freedom, no matter with or without extra auxiliary field(s).

\section{Degeneracy Analysis \label{sec:deg}}

To reduce the number of DOFs, additional conditions must be put on
the Lagrangian (\ref{action_fin}). As we will see in this section,
two conditions are needed in order to make the theory propagate only
two DOFs.

\subsection{The first condition}

As being the first-class constraints, $\pi_{i}$ and $\mathcal{C}_{i}$
correspond to the spatial invariance. Without breaking the spatial
invariance, $\pi_{i}$ and $\mathcal{C}_{i}$ will always be kept
as the first-class constraints, which have vanishing Poisson brackets
with any other constraints. For the sake of simplicity, we are allowed
to omit these two first-class constraints in the subsequent degeneracy
analysis. We thus define a simplified Dirac matrix by omitting the
two columns and rows corresponding to $\pi_{i}$ and $\mathcal{C}_{i}$:
\begin{equation}
\mathcal{M}^{ab}(\vec{x},\vec{y})=\begin{array}{|c|ccc|ccc|}
\hline [\cdot,\cdot] & p(\vec{y}) & \pi(\vec{y}) & p^{kl}(\vec{y}) & \tilde{\pi}^{kl}(\vec{y}) & \mathcal{C}(\vec{y}) & \frac{\delta S}{\delta\phi(\vec{y})}\\
\hline p(\vec{x}) & 0 & 0 & 0 & X & X & X\\
\pi(\vec{x}) & 0 & 0 & 0 & X & X & X\\
p^{ij}(\vec{x}) & 0 & 0 & 0 & X & X & X\\
\hline \tilde{\pi}^{ij}(\vec{x}) & X & X & X & X & X & X\\
\mathcal{C}(\vec{x}) & X & X & X & X & X & X\\
\frac{\delta S}{\delta\phi(\vec{x})} & X & X & X & X & X & 0
\\\hline \end{array}.\label{calM}
\end{equation}

Our purpose is to find conditions to eliminate one scalar type degree
of freedom. Generally we have two choices. One is to introduce extra
conditions to change the second-class constraints into the first-class
constraints, the other is to generate new secondary constraints. In
both cases, a necessary conditions is that the simplified Dirac matrix
$\mathcal{M}^{ab}$ must be degenerate. The determinant of the simplified
Dirac matrix $\mathcal{M}^{ab}$ is
\begin{equation}
\det\mathcal{M}^{ab}(\vec{x},\vec{y})=\det\mathcal{B}^{ab}(\vec{x},\vec{y})\times\det\mathcal{B}^{ba}(\vec{y},\vec{x}),\label{detcalM}
\end{equation}
where the matrix $\mathcal{B}^{ab}$ is the lower left corner of $\mathcal{M}^{ab}$.
The degeneracy of $\mathcal{M}^{ab}$, i.e., $\det\mathcal{M}^{ab}=0$,
thus implies that the sub-matrix $\mathcal{B}^{ab}$ of $\mathcal{M}^{ab}$
is degenerate, i.e., $\det\mathcal{B}^{ab}=0$.

Apparently, the determinant of matrix $\mathcal{B}^{ab}$ is the matrix
form of the first 2DOF condition. This is not the case. Because the
2DOF condition should be an equation that the action satisfies, and
should not contain conjugate momentum. Otherwise, we get the relationship
between the action and the conjugate momentum, which is actually a
constraint. By careful observation, all elements of the matrix $\mathcal{B}^{ab}$
are independent of conjugate momentum, except of $\mathcal{B}^{21}$
given by
\begin{eqnarray}
\mathcal{B}^{21}(\vec{x},\vec{y}) & = & \left[\mathcal{C}(\vec{x}),p^{kl}(\vec{y})\right]\nonumber \\
 & = & 2\delta^{3}(\vec{x}-\vec{y})\pi^{kl}(\vec{x})-\frac{\delta^{2}S}{\delta N(\vec{x})\delta B_{kl}(\vec{y})},\label{calB21}
\end{eqnarray}
which can be transformed into
\begin{equation}
\mathcal{B}^{21}(\vec{x},\vec{y})\approx-\delta^{3}(\vec{x}-\vec{y})\frac{1}{N(\vec{y})}\frac{\delta S}{\delta B_{ij}(\vec{y})}+\frac{\delta^{2}S}{\delta B_{ij}(\vec{x})\delta N(\vec{y})}\label{calB21cs}
\end{equation}
on the constraint surface of $\tilde{\pi}^{ij}$. In subsequent calculations,
we always use (\ref{calB21cs}) as the value of $\mathcal{B}^{21}$
instead of (\ref{calB21}). As a result, the matrix form of the first
2DOF condition is
\begin{equation}
\mathcal{S}_{1}(\vec{x},\vec{y})=\det\mathcal{B}^{ab}(\vec{x},\vec{y})\approx0,\label{calS1}
\end{equation}
where
\begin{equation}
\mathcal{B}^{ab}=\begin{array}{|c|ccc|}
\hline [\cdot,\cdot] & p(\vec{y}) & \pi(\vec{y}) & p^{kl}(\vec{y})\\
\hline -2N(\vec{x})\tilde{\pi}^{ij}(\vec{x}) & \frac{\delta^{2}S}{\delta B_{ij}(\vec{x})\delta\phi(\vec{y})} & N(\vec{x})\frac{\delta}{\delta N(\vec{y})}\left(\frac{1}{N(\vec{x})}\frac{\delta S}{\delta B_{ij}(\vec{x})}\right) & \frac{\delta^{2}S}{\delta B_{ij}(\vec{x})\delta B_{kl}(\vec{y})}\\
-\mathcal{C}(\vec{x}) & \frac{\delta^{2}S}{\delta N(\vec{x})\delta\phi(\vec{y})} & \frac{\delta^{2}S}{\delta N(\vec{x})\delta N(\vec{y})} & N(\vec{y})\frac{\delta}{\delta N(\vec{x})}\left(\frac{1}{N(\vec{y})}\frac{\delta S}{\delta B_{kl}(\vec{y})}\right)\\
\frac{\delta S}{\delta\phi(\vec{x})} & \frac{\delta^{2}S}{\delta\phi(\vec{x})\delta\phi(\vec{y})} & \frac{\delta^{2}S}{\delta\phi(\vec{x})\delta N(\vec{y})} & \frac{\delta^{2}S}{\delta\phi(\vec{x})\delta B_{kl}(\vec{y})}
\\\hline \end{array}.\label{calBab_def}
\end{equation}

In the above, we have adjusted the coefficients of $\tilde{\pi}^{ij}$
and $\mathcal{C}$ in the matrix $\mathcal{B}^{ab}$ to make the it
more symmetric. Since $\mathcal{B}^{ab}$ is a degenerate matrix,
there must be a null eigenvector $\mathcal{V}_{a}$ satisfying
\begin{equation}
\int\mathrm{d}^{3}y\,\mathcal{B}^{ab}(\vec{x},\vec{y})\mathcal{V}_{b}(\vec{y})\approx0^{a}.\label{eqnullev}
\end{equation}
Generally, $\mathcal{V}_{a}$ takes the form
\begin{equation}
\mathcal{V}_{a}(\vec{x})=\left(\begin{array}{ccc}
\left(\mathcal{\mathcal{V}}_{1}\right)_{ij} & \mathcal{\mathcal{V}}_{2} & \mathcal{\mathcal{V}}_{3}\end{array}\right)(\vec{x})=\int\mathrm{d}^{3}y\,U_{a}(\vec{x},\vec{y})V(\vec{y}).\label{nev_calVa}
\end{equation}
In (\ref{nev_calVa}), $V(\vec{x})$ is an arbitrary function of spatial
coordinates and $U_{a}(\vec{x},\vec{y})$ depends on the phase space
variables. One method to get the explicit form of the null eigenvector
$\mathcal{V}_{a}$ is to calculate the adjoint matrix of matrix $\mathcal{B}^{ab}$,
which can be found in Appendix \ref{app:nev}. 

Assuming that $\mathcal{B}_{2}^{*}$ does not vanish, we can use the
linear combination of the null eigenvector $\mathcal{V}_{a}$ and
constraint $\tilde{\pi}^{ij}$, $\mathcal{C}$, $\frac{\delta S}{\delta\phi}$
to define a new constraint $\mathcal{C}'$,
\begin{equation}
\int\mathrm{d}^{3}x\,\mathcal{C}'(\vec{x})V(\vec{x})=\int\mathrm{d}^{3}x\left(\begin{array}{ccc}
\tilde{\pi}^{ij} & \mathcal{C} & \frac{\delta S}{\delta\phi}\end{array}\right)(\vec{x})\left(\begin{array}{ccc}
\left(\mathcal{\mathcal{V}}_{1}\right)_{ij} & \mathcal{\mathcal{V}}_{2} & \mathcal{\mathcal{V}}_{3}\end{array}\right)(\vec{x})\approx0.\label{calCp}
\end{equation}
In (\ref{calCp}), $V(\vec{x})$ is the arbitrary function in (\ref{nev_calVa}).

The merit of introducing the new constraint $\mathcal{C}'$ is that
the Poisson brackets between $\mathcal{C}'$ and constraints $p$,
$\pi$, $p^{kl}$ are all vanishing
\begin{equation}
\left[\mathcal{C}'(\vec{x}),p(\vec{y})\right]\approx\left[\mathcal{C}'(\vec{x}),\pi(\vec{y})\right]\approx\left[\mathcal{C}'(\vec{x}),p^{kl}(\vec{y})\right]\approx0.\label{PBcalCp}
\end{equation}

With this new set of primary constraints, the consistency conditions
of the new constraint $\mathcal{C}'$ reduce to
\begin{equation}
\begin{aligned}\frac{\mathrm{d}\mathcal{C}'(\vec{x})}{\mathrm{d}t} & =\left[\mathcal{C}'(\vec{x})(\vec{x}),H_{\mathrm{T}}\right]\\
 & =\int\mathrm{d}^{3}y\left[\mathcal{C}'(\vec{x}),\tilde{\pi}^{kl}(\vec{y})\right]\lambda_{kl}(\vec{y})+\left[\mathcal{C}'(\vec{x}),H_{\mathrm{C}}\right]\approx0
\end{aligned}
\label{cccalCp}
\end{equation}
where $\lambda_{ij}$ has been fixed to be zero in (\ref{conscond_mat}).
If $\left[\mathcal{C}'(\vec{x}),H_{\mathrm{C}}\right]\neq0$, (\ref{cccalCp})
yields a new secondary constraint $\Phi(\vec{y})$:
\begin{equation}
\Phi(\vec{y})=\left[\mathcal{C}'(\vec{y}),H_{C}\right]\approx0.\label{PhiPB}
\end{equation}

Otherwise, there is no secondary constraint, but we can always find
a new first class constraint through linear combination of constraints
$p$, $\pi$ and $p^{ij}$. Because the necessary and sufficient condition
for generating secondary constraints is that the consistency condition
cannot be automatically satisfied, rather than requiring the simplified
Dirac matrix to be degenerate. Therefore, after requiring the consistency
condition, two cases may arise:
\begin{enumerate}
\item The consistency condition is not automatically satisfied, so that
an additional secondary constraint is required to ensure that the
consistency condition is satisfied;
\item The consistency condition is automatically satisfied, and a first
class constraint is obtained by combining the existing constraints.
\end{enumerate}
When we require a specific 2DOF condition, we should check whether
this condition violates the consistency condition and generates additional
secondary constraints. We should consider this newly generated secondary
constraint in the subsequent discussion.

\subsection{The second condition}

If $\left[\mathcal{C}^{\prime}(\vec{x}),H_{\mathrm{C}}\right]\neq0$
in the previous section, we introduce a new constraint $\pi'$ in
the same way as $\mathcal{C}'$,
\begin{equation}
\int\mathrm{d}^{3}x\,\mathcal{\pi}'(\vec{x})V(\vec{x})\equiv\int\mathrm{d}^{3}x\left(\begin{array}{ccc}
p^{ij} & \pi & p\end{array}\right)(\vec{x})\left(\begin{array}{ccc}
\left(\mathcal{\mathcal{V}}'_{1}\right)_{ij} & \mathcal{\mathcal{V}}'_{2} & \mathcal{\mathcal{V}}'_{3}\end{array}\right)(\vec{x})\approx0,\label{piprime}
\end{equation}
where again $V(\vec{x})$ is the arbitrary function in (\ref{nev_calVa}).
Similar to constraint $\mathcal{C}'$, the new constraint $\mathcal{\pi}'$
has the vanishing Poisson brackets
\begin{equation}
\left[\tilde{\pi}^{ij}(\vec{x}),\mathcal{\pi}'(\vec{y})\right]\approx\left[\mathcal{C}'(\vec{x}),\mathcal{\pi}'(\vec{y})\right]\approx\left[\frac{\delta S}{\delta\phi(\vec{x})},\mathcal{\pi}'(\vec{y})\right]\approx0.\label{PBpiprm}
\end{equation}

By making use of the constraints $\mathcal{\pi}'$, $\mathcal{C}'$
and $\Phi$, the simplified Dirac matrix becomes
\begin{equation}
\mathcal{M}'{}^{ab}=\begin{array}{|c|ccccc|}
\hline [\cdot,\cdot] & \mathcal{\pi}'(\vec{y}) & \left\{ p(\vec{y}),p^{kl}(\vec{y})\right\}  & \mathcal{C}'(\vec{y}) & \left\{ \tilde{\pi}^{kl}(\vec{y}),\frac{\delta S}{\delta\phi(\vec{y})}\right\}  & \Phi(\vec{x})\\
\hline \mathcal{\pi}'(\vec{x}) & 0 & 0 & 0 & 0 & X\\
\left\{ p(\vec{x}),p^{kl}(\vec{x})\right\}  & 0 & 0 & 0 & -\mathcal{C}^{ba}(\vec{y},\vec{x}) & X\\
\mathcal{C}'(\vec{x}) & 0 & 0 & X & X & X\\
\left\{ \tilde{\pi}^{kl}(\vec{x}),\frac{\delta S}{\delta\phi(\vec{x})}\right\}  & 0 & \mathcal{C}^{ab}(\vec{x},\vec{y}) & X & X & X\\
\Phi(\vec{x}) & X & X & X & X & X
\\\hline \end{array}\label{DiracM}
\end{equation}

Since the first 2DOF condition $\mathcal{S}_{1}(\vec{x},\vec{y})$
has been imposed, which yields a secondary constraint to reduced the
dimension of the phase space by one and thus reduce a half degree
of freedom. We need to require an additional 2DOF condition to reduce
the number of degrees of freedom to 2, which means that the simplified
Dirac matrix is required to be degenerate again. The determinant of
the matrix is proportional to $\left[\mathcal{C}'(\vec{x}),\mathcal{C}'(\vec{y})\right]$,
$\left[\Phi(\vec{x}),\mathcal{\pi}'(\vec{y})\right]$ and $\det\mathcal{C}^{ab}(\vec{x},\vec{y})$.
If the determinant of this matrix is required to be vanishing, two
types of 2DOF conditions can be obtained:
\begin{eqnarray}
\mathcal{S}_{2}^{(1)}(\vec{x},\vec{y}) & = & \det\mathcal{C}^{ab}(\vec{x},\vec{y})\times\left[\Phi(\vec{x}),\mathcal{\pi}'(\vec{y})\right]=\det\mathcal{D}^{ab}(\vec{x},\vec{y})\approx0,\label{calS21}\\
\mathcal{S}_{2}^{(2)}(\vec{x},\vec{y}) & = & \left[\mathcal{C}'(\vec{x}),\mathcal{C}'(\vec{y})\right]\approx0,\label{calS22}
\end{eqnarray}
where
\begin{align}
\mathcal{D}^{ab}(\vec{x},\vec{y}) & =\begin{array}{|c|ccc|}
\hline [\cdot,\cdot] & p(\vec{y}) & \pi(\vec{y}) & p^{kl}(\vec{y})\\
\hline \tilde{\pi}^{ij}(\vec{x}) & X & X & X\\
\frac{\delta S}{\delta\phi(\vec{x})} & X & X & X\\
\Phi(\vec{x}) & X & X & X
\\\hline \end{array}.\label{calDab}
\end{align}

The 2DOF condition $\mathcal{S}_{2}^{(1)}$ will generate a secondary
constraint, if the consistency condition is not automatically satisfied.
On the other hand, the 2DOF condition $\mathcal{S}_{2}^{(2)}$ clearly
generates a first-class constraint.

If $\left[\mathcal{C}'(\vec{x}),H_{\mathrm{C}}\right]=0$ in the previous
section, the second 2DOF conditions will become
\begin{eqnarray}
\mathcal{S}'{}_{2}^{(1)}(\vec{x},\vec{y}) & = & \det\mathcal{C}^{ab}(\vec{x},\vec{y})\approx0,\label{calSp21}\\
\mathcal{S}'{}_{2}^{(2)}(\vec{x},\vec{y}) & = & \left[\mathcal{C}'(\vec{x}),\mathcal{C}'(\vec{y})\right]\approx0,\label{calSp22}
\end{eqnarray}
which is covered by TTDOF conditions $\mathcal{S}_{2}^{(1)}$and $\mathcal{S}_{2}^{(2)}$.

The space of theories satisfying the 2DOF conditions $\mathcal{S}_{1}$,
$\mathcal{S}{}_{2}^{(1)}$ and $\mathcal{S}_{2}^{(2)}$ can be divided
into three branches:
\begin{enumerate}
\item Linear constraints branch: If the secondary constraints are linearly
dependent, 2DOF conditions will be strongly vanishing. However, in
this case the number of degrees of freedom will not change. Due to
the linear dependence, the number of secondary constraints will be
reduced by one, which will offset the half degree of freedom reduction
caused by the 2DOF conditions. For example, if constraints $\tilde{\pi}^{ij}$,
$\mathcal{C}$ and $\frac{\delta S}{\delta\phi}$ are linearly dependent,
2DOF condition $\mathcal{S}_{1}$ is automatically satisfied. However,
in this case the number of degrees of freedom remains unchanged.
\item Non-physical branch: Some 2DOF conditions may make the theory be trival.
For example, if the secondary constraints generated by the consistency
condition still cannot satisfy the consistency condition, additional
secondary constraints will be generated. However, these additional
secondary constraints may reduce the number of degrees of freedom
to be less than 2, which is an unphysical situation.
\item Physical branch: In this branch, we get a scalar-tensor theory with
only two tensorial degrees of freedom. 
\end{enumerate}
In the following when we mention 2DOF conditions, we always refer
to the third physical branch.

\section{A concrete model of $d=2$ \label{sec:mod}}

\subsection{The Lagrangian}

In \citep{Gao:2020yzr,Gao:2020qxy} (see also \citep{Gao:2019liu}),
monomials of SCG have been constructed and classified according to
the total number of derivatives $d$ in each monomial. As an example
of our formalism, in this section we will apply the two conditions
to the concrete model of $d=2$. The action of is given by
\begin{equation}
\tilde{S}_{2}\left(\phi,N,h_{ij},{}^{3}\!R_{ij},K_{ij};\mathrm{D}_{i}\right)=S_{2}\left(\phi,N,h_{ij},{}^{3}\!R_{ij},B_{ij};\mathrm{D}_{i}\right)+\int\mathrm{d}t\mathrm{d}^{3}x\frac{\delta S_{2}}{\delta B_{ij}}\left(K_{ij}-B_{ij}\right).\label{action_tm}
\end{equation}

As being discussed before, we introduce the variable $B_{ij}$ in
order to make the explicit reversion of velocity $\dot{h}_{ij}$ in
terms of the momentum possible. The equivalent action $S_{2}$ can
be written as
\begin{equation}
S_{2}=\int\mathrm{d}t\mathrm{d}^{3}x\,N\sqrt{h}\mathcal{L}_{2},\label{S2toy}
\end{equation}
where
\begin{equation}
\mathcal{L}_{2}=c_{1}B_{ij}B^{ij}+c_{2}B^{2}+c_{3}R+c_{4}a_{i}a^{i}+d_{1}\mathrm{D}_{i}\phi\mathrm{D}^{i}\phi+d_{2}a_{i}\mathrm{D}^{i}\phi.\label{Lag_mod}
\end{equation}
The coefficients $c_{i}$ and $d_{i}$ are general functions of $N$
and $\phi$. The acceleration $a_{i}$ is defined as
\begin{equation}
a_{i}=\frac{1}{N}\mathrm{D}_{i}N.\label{acce}
\end{equation}

For later convenience, we can evaluate some of the elements of $\mathcal{B}^{ab}$
as
\begin{eqnarray}
\frac{\delta^{2}S}{\delta B_{ij}(\vec{x})\delta B_{mn}(\vec{y})} & = & 2N\sqrt{h}\left(\frac{1}{2}c_{1}(\delta^{im}\delta^{jn}+\delta^{in}\delta^{jm})+c_{2}\delta^{ij}\delta^{mn}\right)(\vec{x})\delta^{3}(\vec{x}-\vec{y}),\label{calBab1}\\
N(\vec{y})\frac{\delta}{\delta N(\vec{x})}\left(\frac{1}{N(\vec{y})}\frac{\delta S_{B}}{\delta B_{ij}(\vec{y})}\right) & = & 2N\sqrt{h}\left(\frac{\partial c_{1}}{\partial N}B^{ij}+\frac{\partial c_{2}}{\partial N}B\delta^{ij}\right)(\vec{x})\delta^{3}(\vec{x}-\vec{y}),\label{calBab2}\\
\frac{\delta^{2}S}{\delta B_{ij}(\vec{x})\delta\phi(\vec{y})} & = & 2N\sqrt{h}\left(\frac{\delta c_{1}}{\delta\phi}B^{ij}+\frac{\delta c_{2}}{\delta\phi}B\delta^{ij}\right)(\vec{x})\delta^{3}(\vec{x}-\vec{y}).\label{calBab3}
\end{eqnarray}

In order to prevent the dynamical terms of the theory from degenerating,
we assume
\begin{equation}
\det\frac{\delta^{2}S}{\delta B_{ij}(\vec{x})\delta B_{mn}(\vec{y})}=\left(2N\sqrt{h}\delta^{3}(\vec{x}-\vec{y})\right)^{6}c_{1}^{6}(c_{1}+3c_{2})\neq0,\label{detSB2d}
\end{equation}
which implies
\begin{equation}
c_{1}\neq0,\qquad c_{1}+3c_{2}\neq0.\label{conddetSB}
\end{equation}
As long as (\ref{conddetSB}) is satisfied, the non-physical branch
of the 2DOF conditions is naturally removed.

Other elements of matrix $\mathcal{B}^{ab}$ are evaluated to be
\begin{eqnarray}
\frac{\delta^{2}S}{\delta\phi(\vec{x})\delta\phi(\vec{y})} & = & N\sqrt{h}\left[\left(\frac{\delta^{2}S}{\delta\phi\delta\phi}\right)_{0}+\right.\nonumber \\
 &  & \quad\left(-2\frac{\partial d_{1}}{\partial\phi}\mathrm{D}^{i}\phi+\Big(-2N\frac{\partial d_{1}}{\partial N}-2d_{1}\Big)a^{i}\right)\mathrm{D}_{i}\nonumber \\
 &  & \quad\left.-2d_{1}\mathrm{D}^{i}\mathrm{D}_{i}\right](\vec{x})\delta^{3}(\vec{x}-\vec{y}),\label{d2Sff}
\end{eqnarray}
\begin{eqnarray}
\frac{\delta^{2}S}{\delta N(\vec{x})\delta N(\vec{y})} & = & N\sqrt{h}\left[\left(\frac{\delta^{2}S}{\delta N\delta N}\right)_{0}+\right.\nonumber \\
 &  & \quad+\left(-\frac{2}{N^{2}}\frac{\partial c_{4}}{\partial\phi}\mathrm{D}^{i}\phi+\Big(\frac{2c_{4}}{N^{2}}-\frac{2}{N}\frac{\partial c_{4}}{\partial N}\Big)a^{i}\right)\mathrm{D}_{i}\nonumber \\
 &  & \quad\left.-\frac{2c_{4}}{N^{2}}\mathrm{D}^{i}\mathrm{D}_{i}\right](\vec{x})\delta^{3}(\vec{x}-\vec{y}),\label{d2SNN}
\end{eqnarray}
\begin{eqnarray}
\frac{\delta}{\delta\phi(\vec{y})}\left(\frac{\delta S}{\delta N(\vec{x})}\right) & = & N\sqrt{h}\left[\left(\frac{\delta^{2}S}{\delta\phi\delta N}\right)_{0}+\right.\nonumber \\
 &  & \quad\left(\Big(\frac{2d_{1}}{N}+2\frac{\partial d_{1}}{\partial N}-\frac{2}{N}\frac{\partial d_{2}}{\partial\phi}\Big)\mathrm{D}^{i}\phi-\frac{2}{N}\frac{\partial c_{4}}{\partial\phi}a^{i}\right)\mathrm{D}_{i}\nonumber \\
 &  & \quad\left.-\frac{d_{2}}{N}\mathrm{D}^{i}\mathrm{D}_{i}\right](\vec{x})\delta^{3}(\vec{x}-\vec{y}),\label{d2SfN}
\end{eqnarray}
\begin{eqnarray}
\frac{\delta}{\delta N(\vec{y})}\left(\frac{\delta S}{\delta\phi(\vec{x})}\right) & = & N\sqrt{h}\left[\left(\frac{\delta^{2}S}{\delta N\delta\phi}\right)_{0}+\right.\nonumber \\
 &  & \quad\left(\Big(-2\frac{\partial d_{1}}{\partial N}-\frac{2d_{1}}{N}\Big)\mathrm{D}^{i}\phi+\Big(\frac{2}{N}\frac{\partial c_{4}}{\partial\phi}-2\frac{\partial d_{2}}{\partial N}\Big)a^{i}\right)\mathrm{D}_{i}\nonumber \\
 &  & \quad\left.-\frac{d_{2}}{N}\mathrm{D}^{i}\mathrm{D}_{i}\right](\vec{x})\delta^{3}(\vec{x}-\vec{y}),\label{d2SNf}
\end{eqnarray}
where $\left(\frac{\delta^{2}S}{\delta\phi\delta\phi}\right)_{0}$
represents terms in $\frac{\delta^{2}S}{\delta\phi(\vec{x})\delta\phi(\vec{y})}$
that are independent of derivatives of the delta function. We use
this notation because their specific expressions are very complicated
and only the coefficients of derivatives of the delta function are
relevant to the subsequent analysis.

\subsection{The first condition}

We have already obtained the first 2DOF condition in (\ref{calS1}).
To get the specific form of the 2DOF conditions for the this  model,
a straightforward approach is to calculate the determinant of the
matrix $\mathcal{B}^{ab}$, which is however is very complicated and
not illuminating. Alternatively, we can choose an equivalent approach
by assuming that a null eigenvector $\mathcal{V}_{a}(\vec{x})$ exists
and find the expression of this null eigenvector. In this approach,
we can find which conditions are required to make the Dirac matrix
$\mathcal{B}^{ab}$ degenerate.

We assume $U_{a}(\vec{x},\vec{y})$ in (\ref{nev_calVa}) is independent
of the derivative of the delta function. As a result, different orders
of derivative of the delta function will be separated in the matrix
$\mathcal{B}^{ab}$,
\begin{equation}
\mathcal{B}^{ab}(\vec{x},\vec{y})=\mathcal{B}_{(0)}^{ab}(\vec{x},\vec{y})+\mathcal{B}_{(1)}^{ab}(\vec{x},\vec{y})+\mathcal{B}_{(2)}^{ab}(\vec{x},\vec{y}),\label{calBabsep}
\end{equation}
in which terms with different order (of derivative of the delta function)
in the matrix $\mathcal{B}^{ab}$ separately satisfies
\begin{equation}
\int\mathrm{d}^{3}x\,\mathcal{V}_{a}(\vec{x})\mathcal{B}_{(i)}^{ab}(\vec{x},\vec{y})\approx0^{b},\quad i=0,1,2.\label{neveq}
\end{equation}
Note by observing the form of the secondary constraint $\mathcal{C}$
and $\frac{\delta S}{\delta\phi}$, (\ref{neveq}) can only be weakly
satisfied when $i=0$.

In matrix $\mathcal{B}_{(1)}^{ab}(\vec{x},\vec{y})$ and $\mathcal{B}_{(2)}^{ab}(\vec{x},\vec{y})$,
only the $2\times2$ sub-matrices in the lower left corner are not
vanishing, which are
\begin{equation}
\left(\begin{array}{cc}
2\left(\frac{d_{1}}{N}+\frac{\partial d_{1}}{\partial N}-\frac{1}{N}\frac{\partial d_{2}}{\partial\phi}\right)\mathrm{D}^{i}\phi-\frac{2}{N}\frac{\partial c_{4}}{\partial\phi}a^{i} & -\frac{2}{N^{2}}\frac{\partial c_{4}}{\partial\phi}\mathrm{D}^{i}\phi+2\left(\frac{c_{4}}{N^{2}}-\frac{1}{N}\frac{\partial c_{4}}{\partial N}\right)a^{i}\\
-2\frac{\partial d_{1}}{\partial\phi}\mathrm{D}^{i}\phi-2\left(N\frac{\partial d_{1}}{\partial N}+d_{1}\right)a^{i} & -2\left(\frac{\partial d_{1}}{\partial N}+\frac{d_{1}}{N}\right)\mathrm{D}^{i}\phi+\left(\frac{2}{N}\frac{\partial c_{4}}{\partial\phi}-2\frac{\partial d_{2}}{\partial N}\right)a^{i}
\end{array}\right)_{x}\mathrm{D}_{x^{i}}\delta^{3}\left(\vec{x}-\vec{y}\right)\label{calBab1sub}
\end{equation}
and
\begin{equation}
\left(\begin{array}{cc}
-\frac{d_{2}}{N} & -\frac{2c_{4}}{N^{2}}\\
-2d_{1} & -\frac{d_{2}}{N}
\end{array}\right)_{x}\mathrm{D}^{x^{i}}\mathrm{D}_{x_{i}}\delta^{3}\left(\vec{x}-\vec{y}\right),\label{calBab2sub}
\end{equation}
respectively. In the above a subscript ``$x$'' denotes functional
dependence on $\vec{x}$. In the subsequent discussion, we omit $\vec{x}$
for brevity when the dependence on $\vec{x}$ is clear from the context.

According to (\ref{calBab2sub}), for the null eigenvalue equation
of $\mathcal{B}_{(2)}^{ab}$, the part of matrix $\mathcal{B}^{ab}$
is proportional to $\mathrm{D}^{x^{i}}\mathrm{D}_{x^{i}}\delta^{3}(\vec{x}-\vec{y})$,
which yields
\begin{eqnarray}
d_{2}\mathcal{V}_{2}+2Nd_{1}\mathcal{V}_{3} & = & 0,\label{neeqcalB2_1}\\
2c_{4}\mathcal{V}_{2}+Nd_{2}\mathcal{V}_{3} & = & 0.\label{neeqcalB2_2}
\end{eqnarray}

As for the null eigenvalue equation of $\mathcal{B}_{(1)}^{ab}$,
the part of matrix $\mathcal{B}^{ab}$ only related to $\mathrm{D}_{x^{i}}\delta^{3}(\vec{x}-\vec{y})$,
which yields
\begin{eqnarray}
\left(\frac{2d_{1}}{N}+2\frac{\partial d_{1}}{\partial N}-\frac{2}{N}\frac{\partial d_{2}}{\partial\phi}\right)\mathcal{V}_{2}-2\frac{\partial d_{1}}{\partial\phi}\mathcal{V}_{3} & = & 0,\label{neeqcalB1_1}\\
\frac{2}{N}\frac{\partial c_{4}}{\partial\phi}\mathcal{V}_{2}-\left(2N\frac{\partial d_{1}}{\partial N}+2d_{1}\right)\mathcal{V}_{3} & = & 0,\label{neeqcalB1_2}\\
\frac{2}{N^{2}}\frac{\partial c_{4}}{\partial\phi}\mathcal{V}_{2}-\left(2\frac{\partial d_{1}}{\partial N}+\frac{2d_{1}}{N}\right)\mathcal{V}_{3} & = & 0,\label{neeqcalB1_3}\\
\left(\frac{2c_{4}}{N^{2}}-\frac{2}{N}\frac{\partial c_{4}}{\partial N}\right)\mathcal{V}_{2}-\left(\frac{2}{N}\frac{\partial c_{4}}{\partial\phi}-2\frac{\partial d_{2}}{\partial N}\right)\mathcal{V}_{3} & = & 0.\label{neeqcalB1_4}
\end{eqnarray}
There are four equations since the coefficients of $\mathrm{D}^{i}\phi$
and $a^{i}$ need to be proportional to the null eigenvector. We assume
that $d_{1}$, $d_{2}$ and $c_{4}$ are not vanishing. We can simply
choose
\begin{eqnarray}
\mathcal{V}_{2} & = & 2Nd_{1},\label{calV2}\\
\mathcal{V}_{3} & = & -d_{2},\label{calV3}
\end{eqnarray}
and we can get two independent equations
\begin{eqnarray}
d_{2}^{2}-4d_{1}c_{4} & = & 0,\label{indeq1}\\
\left(\frac{d_{1}}{N}+2\frac{\partial d_{1}}{\partial N}-\frac{2}{N}\frac{\partial d_{2}}{\partial\phi}\right)2Nd_{1}+\frac{\partial d_{1}}{\partial\phi}d_{2} & = & 0.\label{indeq2}
\end{eqnarray}
The other two equations are not independent and can be obtained by
combining these two equations.

We have a set of special solutions
\begin{equation}
d_{1}=\frac{A}{2N},\quad d_{2}=B,\quad c_{4}=\frac{8NB^{2}}{A},\label{spesol1}
\end{equation}
where $A$ and $B$ are constants. As we will see, this set of special
solutions will simplify the calculations in the subsequent discussion. 

When we discuss the degeneracy of the matrix $\mathcal{B}_{(0)}^{ab}$,
we need to be careful to identify which parts weakly equal to zero.
To make this clear, we should first discuss how secondary constraints
participate in the weakly equal equation.

Since in the Lagrangian \ref{action_tm} the the velocity term $\dot{h}_{ij}$
(encoded in $K_{ij}$) has been linearized, each canonical momentum
will correspond to a primary constraint. Then all the secondary constraints
can be combined into a form that are independent of canonical momenta.
For example, we make a linear combination of the constraints $\tilde{\pi}^{ij}$
and $\mathcal{C}$,
\begin{equation}
\mathcal{C}_{1}=2\tilde{\pi}^{ij}B_{ij}-\mathcal{C}=\frac{\delta S}{\delta N}-\frac{B_{ij}}{N}\frac{\delta S}{\delta B_{ij}}.\label{calC1}
\end{equation}
By combining constraints $\mathcal{C}_{1}$ and $\frac{\delta S}{\delta\phi}$,
we can obtain a constraint $\mathcal{C}_{2}$,
\begin{eqnarray}
\mathcal{C}_{2} & = & d_{2}\frac{\delta S}{\delta\phi}-2d_{1}N\mathcal{C}_{1}\nonumber \\
 & = & N\sqrt{h}\left[\left(B\frac{\partial c_{1}}{\partial\phi}-A\Big(-\frac{c_{1}}{N}+\frac{\partial c_{1}}{\partial N}\Big)\right)B_{ij}B^{ij}\right.\nonumber \\
 &  & \left.+\left(B\frac{\partial c_{2}}{\partial\phi}-A\Big(-\frac{c_{2}}{N}+\frac{\partial c_{2}}{\partial N}\Big)\right)B^{2}+\left(B\frac{\partial c_{3}}{\partial\phi}-A\Big(\frac{c_{3}}{N}+\frac{\partial c_{3}}{\partial N}\Big)\right)R\right],\label{calC2}
\end{eqnarray}
which does not depend on the special solution (\ref{spesol1}) we
chose before.

Since we have replaced $\mathcal{B}^{21}$ with a term which is independent
of the canonical momentum, (\ref{calS1}) can be equivalently written
as
\begin{equation}
\det\mathcal{B}^{ab}(\vec{x},\vec{y})=\alpha(\vec{x},\vec{y})\mathcal{C}_{2}(\vec{x})+\beta(\vec{x},\vec{y})\frac{\delta S}{\delta\phi(\vec{x})}.\label{detcalBab}
\end{equation}

By solving equations
\begin{equation}
\int\mathrm{d}^{3}x\,\mathcal{V}_{a}(\vec{x})\mathcal{B}_{(0)}^{ab}(\vec{x},\vec{y})\approx0^{b},\qquad b=3,4,...,8,\label{neeq1}
\end{equation}
we obtain
\begin{equation}
\left(\mathcal{\mathcal{V}}_{1}\right)_{ij}=-2d_{1}N\mathcal{V}_{ij}^{(N)}+d_{2}\mathcal{V}_{ij}^{(\phi)},\label{calV1sol}
\end{equation}
where
\begin{eqnarray}
\mathcal{V}_{ij}^{(N)} & = & \frac{1}{c_{1}}\frac{\partial c_{1}}{\partial N}B_{ij}+\left(\frac{1}{c_{1}+3c_{2}}\frac{\partial c_{2}}{\partial N}-\frac{c_{2}}{(c_{1}+3c_{2})c_{1}}\frac{\partial c_{1}}{\partial N}\right)B\delta_{ij},\label{calVijN}\\
\mathcal{V}_{ij}^{(\phi)} & = & \frac{1}{c_{1}}\frac{\partial c_{1}}{\partial\phi}B_{ij}+\left(\frac{1}{c_{1}+3c_{2}}\frac{\partial c_{2}}{\partial\phi}-\frac{c_{2}}{(c_{1}+3c_{2})c_{1}}\frac{\partial c_{1}}{\partial\phi}\right)B\delta_{ij}.\label{calVijf}
\end{eqnarray}

Since we have previously chosen a set of special solutions for $d_{1}$,
$d_{2}$ and $c_{4}$, equation (\ref{neveq}) with $i=0$ can be
written as
\begin{equation}
\int\mathrm{d}^{3}x\,\mathcal{V}_{a}(\vec{x})\mathcal{B}_{(0)}^{ab}(\vec{x},\vec{y})=D_{1}^{b}(\vec{y})B_{ij}(\vec{y})B^{ij}(\vec{y})+D_{2}^{b}(\vec{y})B^{2}(\vec{y})+D_{3}^{b}(\vec{y})R(\vec{y}),\qquad b=1,2.\label{neeq2}
\end{equation}

Since there are no terms such as $a_{i}a^{i}$ in equation (\ref{neeq2}),
the coefficient $\beta(\vec{x},\vec{y})$ is fixed to zero. The above
formula can be equivalently written as
\begin{equation}
\int\mathrm{d}^{3}x\,\mathcal{V}_{a}(\vec{x})\mathcal{B}_{(0)}^{ab}(\vec{x},\vec{y})=\alpha^{b}(\vec{y})\mathcal{C}_{2}(\vec{y}),\qquad b=1,2.\label{neeq3}
\end{equation}

In order to make (\ref{neeq3}) have a solution, the coefficients
of $B_{ij}B^{ij}$, $B^{2}$ and $R$ need to be vanishing. There
will be 6 equations for $c_{1}$, $c_{2}$ and $c_{3}$. We assume
that
\begin{equation}
\frac{c_{1}}{c_{2}}=\mathrm{const}.\label{c1c2ra}
\end{equation}
This assumption will simplify the expression of $D_{2}^{b}$. 

By fixing the coefficients of $B_{ij}B^{ij}$ and $R$ in (\ref{neeq3}),
we obtain 4 differential equations
\begin{eqnarray}
A\left[\frac{1}{N}\frac{\partial c_{1}}{\partial\phi}+\frac{\partial^{2}c_{1}}{\partial N\partial\phi}-\frac{2}{c_{1}}\frac{\partial c_{1}}{\partial N}\frac{\partial c_{1}}{\partial\phi}-\alpha^{1}\left(-\frac{c_{1}}{N}+\frac{\partial c_{1}}{\partial N}\right)\right]-B\left(\frac{\partial^{2}c_{1}}{\partial\phi^{2}}-\frac{2}{c_{1}}\Big(\frac{\partial c_{1}}{\partial\phi}\Big)^{2}-\alpha^{1}\frac{\partial c_{1}}{\partial\phi}\right) & = & 0,\label{diffeq1}\\
A\left[\frac{2}{N}\frac{\partial c_{1}}{\partial N}+\frac{\partial^{2}c_{1}}{\partial N^{2}}-\frac{2}{c_{1}}\frac{\partial c_{1}}{\partial N}\frac{\partial c_{1}}{\partial N}-\alpha^{2}\left(-\frac{c_{1}}{N}+\frac{\partial c_{1}}{\partial N}\right)\right]-B\left(\frac{1}{N}\frac{\partial c_{1}}{\partial\phi}+\frac{\partial^{2}c_{1}}{\partial N\partial\phi}-\frac{2}{c_{1}}\frac{\partial c_{1}}{\partial N}\frac{\partial c_{1}}{\partial\phi}-\alpha^{2}\frac{\partial c_{1}}{\partial\phi}\right) & = & 0,\nonumber \\
A\left[\frac{1}{N}\frac{\partial c_{3}}{\partial\phi}+\frac{\partial^{2}c_{3}}{\partial N\partial\phi}-\alpha^{1}\left(\frac{1}{N}\frac{\partial c_{3}}{\partial\phi}+\frac{\partial^{2}c_{3}}{\partial N\partial\phi}\right)\right]-B\left(\frac{\partial^{2}c_{3}}{\partial\phi^{2}}-\alpha^{1}\frac{\partial c_{3}}{\partial\phi}\right) & = & 0,\nonumber \\
A\left[\frac{2}{N}\frac{\partial c_{3}}{\partial N}+\frac{\partial^{2}c_{3}}{\partial N^{2}}-\alpha^{2}\left(\frac{1}{N}\frac{\partial c_{3}}{\partial\phi}+\frac{\partial^{2}c_{3}}{\partial N\partial\phi}\right)\right]-B\left(\frac{1}{N}\frac{\partial c_{3}}{\partial\phi}+\frac{\partial^{2}c_{3}}{\partial N\partial\phi}-\alpha^{2}\frac{\partial c_{3}}{\partial\phi}\right) & = & 0.\label{diffeq4}
\end{eqnarray}
As long as equations (\ref{c1c2ra}) and (\ref{diffeq1})-(\ref{diffeq4})
are satisfied, the coefficient of $B^{2}$ will be automatically fixed
to be zero. Therefore, we do not have to discuss the differential
equation corresponding to the coefficient of $B^{2}$ here.

We also choose a set of special solutions
\begin{eqnarray}
c_{1} & = & Nf(\phi),\\
c_{2} & = & CNf(\phi),\\
c_{3} & = & \frac{D}{N}g(\phi),
\end{eqnarray}
where the coefficients $C$ and $D$ are both constant.

By plugging the expressions of $c_{1}$ and $c_{3}$ in (\ref{diffeq1})-(\ref{diffeq4}),
we can simplify the four differential equations
\begin{eqnarray}
\frac{\partial^{2}c_{1}}{\partial\phi^{2}}-\frac{2}{c_{1}}\Big(\frac{\partial c_{1}}{\partial\phi}\Big)^{2}-\alpha^{1}\frac{\partial c_{1}}{\partial\phi} & = & 0,\\
\alpha^{2}\frac{\partial c_{1}}{\partial\phi} & = & 0,\\
\frac{\partial^{2}c_{3}}{\partial\phi^{2}}-\alpha^{1}\frac{\partial c_{3}}{\partial\phi} & = & 0,\\
\alpha^{2}\frac{\partial c_{3}}{\partial\phi} & = & 0.
\end{eqnarray}

Finally, constraint $\mathcal{C}_{2}$ becomes
\begin{align}
\mathcal{C}_{2} & =BN\sqrt{h}\left(\frac{\partial c_{1}}{\partial\phi}B_{ij}B^{ij}+\frac{\partial c_{2}}{\partial\phi}B^{2}+\frac{\partial c_{3}}{\partial\phi}R\right).\label{calC2fin}
\end{align}

As we can see, if both $c_{1}$ and $c_{3}$ are independent of $\phi$,
constraint $\mathcal{C}_{2}$ will do not exist. The number of second-class
constraints will be reduced by one, and the number of degrees of freedom
will be increased by 1/2. This is the linear constraints branch of
2DOF condition we mentioned in the previous section, which is not
the case we are interested in. Therefore we assume that $c_{1}$ and
$c_{3}$ cannot be independent of $\phi$ simultaneously. As a result,
the coefficient $\alpha^{2}$ has to be fixed to be zero. 

\subsection{The second condition}

According to the above discussion, after requiring the first 2DOF
condition $\mathcal{S}_{1}$, we can obtain constraint $\mathcal{C}'$
by the null eigenvector $\mathcal{V}_{a}(\vec{x})$,
\begin{equation}
\int\mathrm{d}^{3}x\,\mathcal{C}'(\vec{x})V(\vec{x})\equiv\int\mathrm{d}^{3}x\left(\begin{array}{ccc}
-2N\tilde{\pi}^{ij} & -\mathcal{C} & \frac{\delta S}{\delta\phi}\end{array}\right)(\vec{x})\left(\begin{array}{ccc}
\left(\mathcal{\mathcal{V}}_{1}\right)_{ij} & \mathcal{\mathcal{V}}_{2} & \mathcal{\mathcal{V}}_{3}\end{array}\right)(\vec{x}),
\end{equation}
which yields
\begin{equation}
\mathcal{C}'=BN\left[\sqrt{h}\left(\frac{\partial c_{1}}{\partial\phi}B^{ij}B_{ij}+\Big(\frac{2c_{2}}{c_{1}}\frac{\partial c_{1}}{\partial\phi}-\frac{\partial c_{2}}{\partial\phi}\Big)B^{2}-\frac{\partial c_{3}}{\partial\phi}R\right)-\frac{2}{c_{1}}\frac{\partial c_{1}}{\partial\phi}\pi^{ij}B_{ij}\right].\label{calCpxpl}
\end{equation}

Poisson brackets between the constraint $\mathcal{C}'$ and constraints
$p$, $\pi$, $p^{kl}$ are
\begin{align}
\left[\mathcal{C}'(\vec{x}),p(\vec{y})\right] & =\left[-BN\frac{\partial}{\partial\phi}\left(\frac{2}{c_{1}}\frac{\partial c_{1}}{\partial\phi}\right)\tilde{\pi}^{ij}B_{ij}-\frac{\partial c_{3}}{\partial\phi\partial\phi}\left(\frac{\partial c_{3}}{\partial\phi}\right)^{-1}\mathcal{C}_{2}\right](\vec{x})\delta^{3}(\vec{x}-\vec{y})\approx0,\label{PBcalCp1}
\end{align}
\begin{equation}
\left[\mathcal{C}'(\vec{x}),\pi(\vec{y})\right]=\frac{B}{N(\vec{x})}\left(\mathcal{C}'(\vec{x})+\mathcal{C}_{2}(\vec{x})\right)\delta^{3}(\vec{x}-\vec{y})\approx0,\label{PBcalCp2}
\end{equation}
\begin{align}
\left[\mathcal{C}'(\vec{x}),p^{kl}(\vec{y})\right] & =-\frac{BN(\vec{x})}{c_{1}(\vec{x})}\frac{\partial c_{1}(\vec{x})}{\partial\phi(\vec{x})}\tilde{\pi}^{kl}(\vec{x})\delta^{3}(\vec{x}-\vec{y})\approx0,\label{PBcalCp3}
\end{align}
which all vanish on the constraint surface. Poisson bracket between
constraint $\mathcal{C}'$ and itself is generally nonzero, unless
\begin{equation}
\frac{\partial c_{1}}{\partial\phi}=0,\quad\text{or}\quad\frac{\partial c_{3}}{\partial\phi}=0.\label{PBcalCp4}
\end{equation}

As we have mentioned above, $c_{1}$ and $c_{3}$ cannot be independent
of $\phi$ simultaneously, otherwise the number of degrees of freedom
will increased by $1/2$. If we require the second 2DOF condition
$\mathcal{S}_{2}^{(2)}$ to be valid, there will be two cases.

In the first case, if we choose $\frac{\partial c_{1}}{\partial\phi}=0$,
constraint $\mathcal{C}_{2}$ will become
\begin{equation}
\mathcal{C}_{2}=BN\sqrt{h}\frac{\partial c_{3}}{\partial\phi}R\approx0.\label{calC2_1}
\end{equation}
This case is clearly non-physical as it fixes $R$ to be vanishing.

On the other hand, if we choose $\frac{\partial c_{3}}{\partial\phi}=0$,
constraint $\mathcal{C}_{2}$ will become
\begin{align}
\mathcal{C}_{2} & =BN\sqrt{h}\left(\frac{\partial c_{1}}{\partial\phi}B_{ij}B^{ij}+\frac{\partial c_{2}}{\partial\phi}B^{2}\right)\approx0.\label{calC2_2}
\end{align}
In this case, we can solve all the coefficients
\begin{align}
c_{1} & =Nf(\phi),\quad c_{2}=CNf(\phi),\quad c_{3}=\frac{D}{N},\nonumber \\
c_{4} & =\frac{8NB^{2}}{A},\quad d_{1}=\frac{A}{2N},\quad d_{2}=B.\label{coeffsol}
\end{align}
The resulting theory thus corresponds to a SCG with an auxiliary scalar
field of $d=2$ that propagates two degrees of freedom. The constraint
coefficients $\alpha^{i}$ in the previous differential equation are
also uniquely determined
\begin{equation}
\alpha^{1}=-\left(\frac{\partial f(\phi)}{\partial\phi}\right)^{-1}\left(\frac{2}{f(\phi)}\Big(\frac{\partial f(\phi)}{\partial\phi}\Big)^{2}-\frac{\partial^{2}f(\phi)}{\partial\phi^{2}}\right),\quad\alpha^{2}=0.\label{alphai}
\end{equation}

Now let us consider the sub-matrix $\mathcal{E}^{ab}$ of the simplified
Dirac matrix $\mathcal{M}'{}^{ab}$,
\begin{equation}
\mathcal{E}^{ab}(\vec{x},\vec{y})=\begin{array}{|c|cccc|}
\hline [\cdot,\cdot] & p(\vec{y}) & \pi(\vec{y}) & p^{kl}(\vec{y}) & \mathcal{C}'(\vec{y})\\
\hline \mathcal{C}'(\vec{x}) & 0 & 0 & 0 & 0\\
\tilde{\pi}^{ij}(\vec{x}) & X & X & X & X\\
\frac{\delta S}{\delta\phi(\vec{x})} & X & X & X & X\\
\Phi(\vec{x}) & X & X & X & X
\\\hline \end{array},\label{calEab}
\end{equation}
whose determinant is obviously zero. Based on the discussion on the
null eigenvectors in Appendix (\ref{app:nev}), we can find a null
eigenvector $\mathcal{V}_{a}^{(\mathcal{E})}$ to build a new first-class
constraint $\tilde{\mathcal{C}}$,
\begin{equation}
\int\mathrm{d}^{3}x\,\tilde{\mathcal{C}}(\vec{x})V(\vec{x})=\int\mathrm{d}^{3}x\left(\begin{array}{cccc}
p & \pi & p^{kl} & \mathcal{C}'\end{array}\right)(\vec{x})\left(\begin{array}{ccc}
\left(\mathcal{\mathcal{V}}{}_{1}\right)_{ij}^{(\mathcal{E})} & \mathcal{\mathcal{V}}{}_{2}^{(\mathcal{E})} & \mathcal{\mathcal{V}}{}_{3}^{(\mathcal{E})}\end{array}\right)(\vec{x})\approx0.\label{tldcalC}
\end{equation}

To summarize, there are 24 constraints which can be divided into two
classes:
\[
\begin{array}{rl}
7\text{ first-class: } & \pi_{i},\quad\mathcal{C}_{i},\quad\tilde{\mathcal{C}},\\
16\text{ second-class: } & p,\quad p^{ij},\quad\pi,\quad\tilde{\pi}^{ij},\quad\frac{\delta S}{\delta\phi},\quad\Phi.
\end{array}
\]
The number of DOFs is thus
\begin{equation}
\begin{aligned}\#_{\text{dof }} & =\frac{1}{2}\left(2\times\#_{\mathrm{var}}-2\times\#_{1\mathrm{st}}-\#_{2\mathrm{nd}}\right)\\
 & =\frac{1}{2}(2\times17-2\times7-16)\\
 & =2.
\end{aligned}
\label{NoDOF}
\end{equation}

Finally let us discuss another 2DOF condition $\mathcal{S}_{2}^{(1)}$.
By applying the consistency condition of constraint $\mathcal{C}_{2}$,
we can obtain the constraint $\Phi$,
\begin{eqnarray}
\Phi(\vec{x}) & = & \left[\mathcal{C}'(\vec{x}),H_{C}\right]\nonumber \\
 & = & -B\frac{2N(\vec{x})}{c_{1}(\vec{x})}\frac{\partial c_{1}(\vec{x})}{\partial\phi(\vec{x})}B_{ij}(\vec{x})\frac{\delta S}{\delta h_{ij}(\vec{x})}\nonumber \\
 &  & -\int d^{3}zBN(\vec{z})\sqrt{h}\frac{\delta}{\delta h_{ij}(\vec{z})}\left(\frac{\partial c_{1}}{\partial\phi}B^{ij}B_{ij}+\Big(\frac{2c_{2}}{c_{1}}\frac{\partial c_{1}}{\partial\phi}-\frac{\partial c_{2}}{\partial\phi}\Big)B^{2}-\frac{\partial c_{3}}{\partial\phi}R\right)(\vec{x})2N(\vec{z})B_{ij}(\vec{z}).\label{Phi}
\end{eqnarray}
Since the second-order derivative of the delta function arises when
calculating the integral in the last line, the constraint $\Phi$
contains $\mathrm{D}^{i}\mathrm{D}_{j}B_{kl}$, which is proportional
to
\begin{equation}
\frac{\partial c_{3}}{\partial\phi}\mathrm{D}^{i}\mathrm{D}_{j}B_{kl}.\label{Phi_DDB}
\end{equation}
So when we require the 2DOF condition $\mathcal{S}_{2}^{(1)}=\det\mathcal{D}^{ab}\approx0$,
it will also imply that $\frac{\partial c_{3}}{\partial\phi}$ may
be vanishing as one of the conditions for the matrix $\mathcal{D}^{ab}$
to be degenerate, which is sufficient to satisfy 2DOF condition $\mathcal{S}_{2}^{(2)}$.

\section{Conclusion \label{sec:con}}

In this work, we investigated a class of spatially covariant gravity
theories with a non-dynamical scalar field. We briefly describe our
model in Sec. \ref{sec:scg}. The general Lagrangian is given by (\ref{actionK}),
which has been proved to have two tensorial and one scalar DOFs \citep{Gao:2018izs}.
The purpose of this work is to determine the conditions on the Lagrangian,
which we dub the 2DOF conditions, under which only two degrees of
freedom are propagating.

A perturbative approach has been taken in \citep{Wang:2024hfd} to
derive conditions such that the scalar mode is eliminated at linear
order in perturbations. In this work we employ the strict Hamiltonian
constraint analysis to derive the conditions such that that only two
DOFs are propagating in the nonperturbative sense. In Sec. \ref{sec:ham},
we describe the Hamiltonian formalism for the primary constraints
and their consistency conditions. As expected, there are 3 DOFs in
the theory if no further conditions are imposed. 

In Sec. \ref{sec:deg}, by requiring the degeneracy of the Dirac matrix,
we find that two conditions are required to make the theory propagate
only 2DOFs. The first condition (\ref{calS1}) states that the matrix
$\mathcal{B}^{ab}$ given in (\ref{calBab_def}) must be degenerate,
which will result in a secondary constraint. In order to fully eliminate
the scalar mode, a second condition is necessary. The second condition
can be divided into two categories according to their different effects
on the Dirac matrix. In the first case, the condition (\ref{calS21})
will generally turn a second-class constraint into a first-class constraint.
In the second case, the condition (\ref{calS22}) will generate another
secondary constraint $\Phi$.

We would like to emphasize that even the Lagrangian satisfies the
both two 2DOF conditions, the number of DOFs do not necessarily decrease.
A special case is that the constraints are linearly dependent, which
implies that the number of DOFs may remain unchanged. Another special
case, which we refer to as the non-physical branch of the 2DOF conditions,
will make the theory trivial or physically unacceptable (e.g., without
tensor modes). One thus must be careful when dealing with the 2DOF
conditions in order to pick up the physical case, i.e., theories with
precisely two degrees of freedom.

In Sec. \ref{sec:mod}, based on the spatially covariant monomials
classified in \citep{Gao:2020yzr,Gao:2020qxy}, we consider
a concrete Lagrangian (\ref{Lag_mod}) built of monomials of $d=2$
(i.e., with two derivatives) as an illustration of our formal analysis.
Since the Dirac matrix of this model contains derivatives of the delta
function, it is complicated to calculate the matrix form of the 2DOF
conditions of this concrete model. Instead, we choose an equivalent
but more efficient way to obtain the 2DOF conditions, which is to
find out the null eigenvector $\mathcal{V}_{a}$ satisfing (\ref{eqnullev}).
The resulting 2DOF conditions are given in (\ref{conddetSB}) and
(\ref{PBcalCp4}). In particular, for this concrete model, we are
able to eliminate the non-physical branch and the linear constraints
branch of the 2DOF conditions. Finally, we find a set of coefficients
(\ref{coeffsol}), and the resulting Lagrangian corresponds to a spatially
covariant gravity theory with an auxiliary scalar field which propagates
only two DOFs.
\begin{acknowledgments}
We would like to thank Zhi-Chao Wang for valuable discussions. X.G.
is supported by the National Natural Science Foundation of China (NSFC)
under Grants No. 12475068 and No. 11975020 and the Guangdong Basic
and Applied Basic Research Foundation under Grant No. 2025A1515012977.
\end{acknowledgments}

\appendix

\section{Comparison with scalar-tensor theory in the spatial gauge \label{app:comp}}

The spatially covariant gravity the an auxiliary scalar field was
firstly proposed in \citep{Gao:2018izs}, in which the Hamiltonian
analysis was performed in order to show that the theory propagates
3 degrees of freedom. This idea was originally motivated by generally
covariant scalar-tensor theory when the scalar field possesses a spacelike
gradient. After choosing the so-called ``spatial gauge'' as in \citep{Gao:2018izs},
the resulting action takes the form that is similar to (\ref{action}).
However, we should notice that ``spatially covariant gravity the
an auxiliary scalar field'' and ``scalar-tensor theory in the spatial
gauge'' are completely different theories. 

First we will show the difference between (\ref{action}) and the
scalar-tensor theory in the spatial gauge. Let us consider a general
action of scalar-tensor theory
\begin{equation}
S_{\mathrm{GST}}=\int\mathrm{d}^{4}x\sqrt{-g}\mathcal{L}\left(\phi;g_{ab},\varepsilon_{abcd},{}^{4}\!R_{abcd};\nabla_{a}\right),\label{S_GST}
\end{equation}
where the Lagrangian involves a scalar field $\phi$, spacetime metric
$g_{ab}$, the spacetime curvature tensor $^{4}\!R_{abcd}$ as well
as their covariant derivatives. The 4-dimension Levi-Civita tensor
$\varepsilon_{abcd}$ encodes possible parity violation effects.

In cosmological context, the scalar field $\phi$ is assumed to possesses
a timelike gradient so that the so-called unitary gauge with $\phi=t$
can be chosen. Here we consider the contrary situation by assuming
that the scalar field possesses a spacelike gradient\footnote{Similar ideas appear in ``elastic inflation'' \citep{Gruzinov:2004ty}
and \textquotedblleft solid inflation\textquotedblright{} \citep{Endlich:2012pz}
where it is the Nambu-Goldstone boson breaking spatial diffeomorphism
that plays the role of the scalar field.}. Contrary to the unitary gauge which is defined by requiring $\mathrm{D}_{a}\phi=0$,
we now can choose a gauge in which
\begin{equation}
\pounds_{\bm{n}}\phi=0,\label{LieDphi0}
\end{equation}
where $n^{a}$ is the normal vector to the hypersurfaces. This can
be understood that we choose hypersurfaces such that the normal vector
$n^{a}$ exactly lie on the the constant $\phi$ hypersurfaces. As
a result, the value of $\phi$ does not change when being transported
along the normal vector. This choice of spatial hypersurfaces is referred
as ``spatial gauge'' in \citep{Gao:2018izs}.

The normal vector $n^{a}$ is normalized by $n^{a}n_{a}=-1$. The
induced metric on the spatial hypersurfaces is as usual
\begin{equation}
h_{ab}=g_{ab}+n_{a}n_{b}.\label{indmet}
\end{equation}
We can then split all the 4-dimensional objects into their temporal
and spatial parts. For example, the decomposition of $\nabla_{a}\phi$
is
\begin{equation}
\nabla_{a}\phi=-n_{a}\pounds_{\boldsymbol{n}}\phi+\mathrm{D}_{a}\phi\xlongequal{\text{spatial gauge}}\mathrm{D}_{a}\phi,\label{dec_nablaphi}
\end{equation}
where $\mathrm{D}_{a}$ is the covariant derivative compatible with
induced metric $h_{ab}$. In the last equality of (\ref{dec_nablaphi})
we have used the fact that $\pounds_{\boldsymbol{n}}\phi=0$ in the
spatial gauge. This is exactly contrary to what in the unitary gauge
where $\mathrm{D}_{a}\phi=0$ and thus $\nabla_{a}\phi\rightarrow-n_{a}\pounds_{\boldsymbol{n}}\phi$.

After making the 3+1 decomposition and choosing the spatial gauge,
the action (\ref{S_GST}) can written in the form
\begin{equation}
S_{\mathrm{GST}}^{\mathrm{(s.g.)}}=\int\mathrm{d}t\mathrm{d}^{3}xN\sqrt{h}\mathcal{L}\left(\phi,N,h_{ij},{}^{3}\!R_{ij};\mathrm{D}_{i},\pounds_{\bm{n}}\right),\label{GSTsg}
\end{equation}
where ``s.g.'' stands for the spatial gauge and we have fixed the
spatial coordinates adapted to the spacelike hypersurfaces. At the
first glance, the action (\ref{GSTsg}) takes the same form as (\ref{action}),
however, the crucial difference is that in the spatial gauge, the
scalar field $\phi$ can be viewed as a time-independent but space-dependent
field, which breaks spatial diffeomorphism. In other words, $\phi=\phi(\bm{x})$
in (\ref{GSTsg}) cannot be viewed as a dynamical nor auxiliary variable,
which is actually a ``function'' of space coordinates with fixed
values. This is completely different from (\ref{action}) which has
spatial covariance, in which $\phi$ plays the role of an auxiliary
field. 

Then we show that the generally covariant version (i.e., correspondence)
of (\ref{action}) is nothing but a ``two-field'' scalar-tensor
theory. Let us take $\mathrm{D}_{i}\phi\mathrm{D}^{i}\phi$ as an
example. The covariant version is
\begin{equation}
\mathrm{D}_{i}\phi\mathrm{D}^{i}\phi=h^{ij}\mathrm{D}_{i}\phi\mathrm{D}_{j}\phi\rightarrow\left(g^{ab}+u^{a}u^{b}\right)\nabla_{a}\phi\nabla_{b}\phi,\label{DfDf}
\end{equation}
where $u_{a}=-N\nabla_{a}\Phi$ with $\Phi$ is the scalar field defining
the spacelike hypersurfaces. In other words, the spacelike hypersurfaces
are defined by hypersurfaces with $\Phi=\mathrm{const.}$. In (\ref{DfDf}),
$N=1/\sqrt{-\left(\nabla\Phi\right)^{2}}$, which reduces to the lapse
function when fixing the so-called unitary gauge with $\Phi=t$. By
expanding (\ref{DfDf}) explicitly, we get
\begin{equation}
\mathrm{D}_{i}\phi\mathrm{D}^{i}\phi\rightarrow\left(g^{ab}-\frac{\nabla^{a}\Phi N\nabla^{b}\Phi}{\left(\nabla\Phi\right)^{2}}\right)\nabla_{a}\phi\nabla_{b}\phi,
\end{equation}
which is clearly a two-field scalar-tensor theory term.

\section{Constraint algebra \label{app:cons}}

In this appendix, we show the explicit expression for the Poisson
brackets. 

The Poisson brackets among constraints are
\begin{eqnarray}
\left[p(\vec{x}),\tilde{\pi}^{kl}(\vec{y})\right] & = & \frac{1}{2N(\vec{y})}\frac{\delta^{2}S}{\delta\phi(\vec{x})\delta B_{kl}(\vec{y})},\\
\left[p^{ij}(\vec{x}),\tilde{\pi}^{kl}(\vec{y})\right] & = & \frac{1}{2}\frac{1}{N(\vec{y})}\frac{\delta^{2}S}{\delta B_{ij}(\vec{x})\delta B_{kl}(\vec{y})},\label{PB_pijpitkl}\\
\left[\pi(\vec{x}),\tilde{\pi}^{ij}(\vec{y})\right] & = & -\frac{1}{2}\delta^{3}(\vec{x}-\vec{y})\frac{1}{N^{2}(\vec{y})}\frac{\delta S}{\delta B_{ij}(\vec{y})}+\frac{1}{2}\frac{1}{N(\vec{y})}\frac{\delta^{2}S}{\delta N(\vec{x})\delta B_{ij}(\vec{y})},\\
\left[\tilde{\pi}^{ij}(\vec{x}),\tilde{\pi}^{kl}(\vec{y})\right] & = & \frac{1}{2N(\vec{y})}\frac{\delta^{2}S}{\delta h_{ij}(\vec{x})\delta B_{kl}(\vec{y})}-\frac{1}{2N(\vec{x})}\frac{\delta^{2}S}{\delta B_{ij}(\vec{x})\delta h_{kl}(\vec{y})},
\end{eqnarray}
\begin{eqnarray}
[\pi(\vec{x}),\mathcal{C}(\vec{y})] & = & \frac{\delta^{2}S}{\delta N(\vec{x})\delta N(\vec{y})},\\{}
[p(\vec{x}),\mathcal{C}(\vec{y})] & = & \frac{\delta^{2}S}{\delta\phi(\vec{x})\delta N(\vec{y})},\\
\left[p^{ij}(\vec{x}),\mathcal{C}(\vec{y})\right] & = & -2\delta^{3}(\vec{x}-\vec{y})\pi^{ij}(\vec{y})+\frac{\delta^{2}S}{\delta B_{ij}(\vec{x})\delta N(\vec{y})},\\
\left[\tilde{\pi}^{ij}(\vec{x}),\mathcal{C}(\vec{y})\right] & = & \frac{\delta^{2}S}{\delta h_{ij}(\vec{x})\delta N(\vec{y})}-\frac{1}{N(\vec{x})}\frac{\delta^{2}S}{\delta B_{ij}(\vec{x})\delta h_{kl}(\vec{y})}B_{kl}(\vec{y}),
\end{eqnarray}
\begin{eqnarray}
\left[p(\vec{x}),\frac{\delta S}{\delta\phi(\vec{y})}\right] & = & -\frac{\delta^{2}S}{\delta\phi(\vec{x})\delta\phi(\vec{y})},\\
\left[p^{ij}(\vec{x}),\frac{\delta S}{\delta\phi(\vec{y})}\right] & = & -\frac{\delta^{2}S}{\delta B_{ij}(\vec{x})\delta\phi(\vec{y})},\\
\left[\pi(\vec{x}),\frac{\delta S}{\delta\phi(\vec{y})}\right] & = & -\frac{\delta^{2}S}{\delta N(\vec{x})\delta\phi(\vec{y})},\\
\left[\tilde{\pi}^{ij}(\vec{x}),\frac{\delta S}{\delta\phi(\vec{y})}\right] & = & -\frac{\delta^{2}S}{\delta h_{ij}(\vec{x})\delta\phi(\vec{y})},
\end{eqnarray}
\begin{eqnarray}
[\mathcal{C}(\vec{x}),\mathcal{C}(\vec{y})] & = & -\frac{\delta^{2}S}{\delta h_{ij}(\vec{y})\delta N(\vec{x})}2B_{ij}(\vec{y})+2B_{ij}(\vec{x})\frac{\delta^{2}S}{\delta h_{ij}(\vec{x})\delta N(\vec{y})},\\
\left[\frac{\delta S}{\delta\phi(\vec{x})},\mathcal{C}(\vec{y})\right] & = & \frac{\delta^{2}S}{\delta\phi(\vec{x})\delta h_{ij}(\vec{y})}2B_{ij}(\vec{y}).
\end{eqnarray}

The Poisson brackets involving the canonical Hamiltonian are
\begin{eqnarray}
\left[\tilde{\pi}^{ij}(\vec{x}),H_{\mathrm{C}}\right] & = & \frac{\delta S}{\delta h_{ij}(\vec{x})}-\frac{1}{N(\vec{x})}\int\mathrm{d}^{3}z\frac{\delta^{2}S}{\delta B_{ij}(\vec{x})\delta h_{kl}(\vec{z})}N(\vec{z})B_{kl}(\vec{z}),\\
\left[\mathcal{C}(\vec{x}),H_{\mathrm{C}}\right] & = & 2B_{ij}(\vec{x})\frac{\delta S}{\delta h_{ij}(\vec{x})}-\int\mathrm{d}^{3}z\frac{\delta S}{\delta N(\vec{x})\delta h_{ij}(\vec{z})}N(\vec{z})2B_{ij}(\vec{z}),\\
\left[\frac{\delta S}{\delta\phi(\vec{x})},H_{\mathrm{C}}\right] & = & \int\mathrm{d}^{3}z\frac{\delta^{2}S}{\delta\phi(\vec{x})\delta h_{ij}(\vec{z})}N(\vec{z})2B_{ij}(\vec{z}).
\end{eqnarray}

\section{Adjoint matrix form of null eigenvector \label{app:nev}}

If we wish to find the null eigenvector of the degenerate matrix $\mathcal{B}^{ab}$,
one method is to calculate the non-zero rows (or columns) of its corresponding
adjoint matrix $\mathcal{B}_{ab}^{*}$, and to multiply the corresponding
elements by $(-1)^{a+b}$ to build the null eigenvector. For simplicity,
we assume that $\mathcal{B}^{ab}$ is a simple $3\times3$ matrix
and the rows and columns correspond to constraints $\phi_{1}$, $\phi_{2}$
and $\phi_{3}$.

To verify that this form of null eigenvector is feasible, we assume
that the matrix $\mathcal{B}^{ab}$ is degenerate and without loss
of generality we take the first row of its adjoint matrix $\mathcal{B}_{ab}^{*}$
to form a null eigenvector
\begin{equation}
\mathcal{V}_{a}=\left(\begin{array}{ccc}
\mathcal{B}_{11}^{*} & -\mathcal{B}_{21}^{*} & \mathcal{B}_{31}^{*}\end{array}\right).
\end{equation}
We then multiply the null eigenvector $\mathcal{V}_{a}$ by the matrix
$\mathcal{B}_{ab}^{*}$,
\begin{equation}
\mathcal{V}_{a}\mathcal{B}^{ab}=\left(\begin{array}{ccc}
\mathcal{B}_{11}^{*} & -\mathcal{B}_{21}^{*} & \mathcal{B}_{31}^{*}\end{array}\right)\left(\begin{array}{ccc}
\mathcal{B}^{1b} & \mathcal{B}^{2b} & \mathcal{B}^{3b}\end{array}\right)^{T},
\end{equation}
which yields
\begin{eqnarray}
\mathcal{V}_{a}\mathcal{B}^{a1} & = & \left(\begin{array}{ccc}
\mathcal{B}_{11}^{*} & -\mathcal{B}_{21}^{*} & \mathcal{B}_{31}^{*}\end{array}\right)\left(\begin{array}{ccc}
\mathcal{B}^{11} & \mathcal{B}^{21} & \mathcal{B}^{31}\end{array}\right)^{T}\nonumber \\
 & = & \det\mathcal{B}^{ab}=0,
\end{eqnarray}
\begin{eqnarray}
\mathcal{V}_{a}\mathcal{B}^{a2} & = & \left(\begin{array}{ccc}
\mathcal{B}_{11}^{*} & -\mathcal{B}_{21}^{*} & \mathcal{B}_{31}^{*}\end{array}\right)\left(\begin{array}{ccc}
\mathcal{B}^{12} & \mathcal{B}^{22} & \mathcal{B}^{32}\end{array}\right)^{T}\nonumber \\
 & = & \det\left(\begin{array}{ccc}
\mathcal{B}^{12} & \mathcal{B}^{12} & \mathcal{B}^{13}\\
\mathcal{B}^{22} & \mathcal{B}^{22} & \mathcal{B}^{23}\\
\mathcal{B}^{32} & \mathcal{B}^{32} & \mathcal{B}^{33}
\end{array}\right)=0,
\end{eqnarray}
\begin{align}
\mathcal{V}_{a}\mathcal{B}^{a3} & =\left(\begin{array}{ccc}
\mathcal{B}_{11}^{*} & -\mathcal{B}_{21}^{*} & \mathcal{B}_{31}^{*}\end{array}\right)\left(\begin{array}{ccc}
\mathcal{B}^{13} & \mathcal{B}^{23} & \mathcal{B}^{33}\end{array}\right)^{T}\nonumber \\
 & =\det\left(\begin{array}{ccc}
\mathcal{B}^{13} & \mathcal{B}^{12} & \mathcal{B}^{13}\\
\mathcal{B}^{23} & \mathcal{B}^{22} & \mathcal{B}^{23}\\
\mathcal{B}^{33} & \mathcal{B}^{32} & \mathcal{B}^{33}
\end{array}\right)=0.
\end{align}

We can combine constraints $\phi_{i}$ with the null eigenvectors
to make a new constraint
\begin{equation}
\phi'(\vec{y})=\int\mathrm{d}^{3}x\left(\begin{array}{ccc}
\phi_{1} & \phi_{2} & \phi_{3}\end{array}\right)(\vec{x})\left(\begin{array}{ccc}
\mathcal{B}_{11}^{*} & -\mathcal{B}_{21}^{*} & \mathcal{B}_{31}^{*}\end{array}\right)^{T}(\vec{x},\vec{y}).
\end{equation}
The Poisson brackets of $\phi'$ with constraints $\phi_{i}$ are
all vanishing,
\begin{eqnarray}
\left[\phi_{i}(\vec{x}),\phi'(\vec{y})\right] & \approx & \int\mathrm{d}^{3}z\left[\phi_{i}(\vec{x}),\left(\begin{array}{ccc}
\phi_{1} & \phi_{2} & \phi_{3}\end{array}\right)(\vec{z})\right]\left(\begin{array}{ccc}
\mathcal{B}_{11}^{*} & -\mathcal{B}_{21}^{*} & \mathcal{B}_{31}^{*}\end{array}\right)^{T}(\vec{z},\vec{y})\nonumber \\
 & = & \int\mathrm{d}^{3}z\left(\begin{array}{ccc}
\mathcal{B}^{1i} & \mathcal{B}^{2i} & \mathcal{B}^{3i}\end{array}\right)\left(\begin{array}{ccc}
\mathcal{B}_{11}^{*} & -\mathcal{B}_{21}^{*} & \mathcal{B}_{31}^{*}\end{array}\right)^{T}(\vec{z},\vec{y})\nonumber \\
 & = & 0.
\end{eqnarray}

In practical calculations, in order to avoid integrating the delta
function of higher order, we need to multiply the null eigenvector
by a greatest common divisor, i.e.,
\begin{equation}
\mathcal{V}_{a}=\left(\begin{array}{ccc}
\mathcal{\mathcal{V}}_{1} & \mathcal{\mathcal{V}}_{2} & \mathcal{\mathcal{V}}_{3}\end{array}\right)=\frac{1}{\mathrm{gcd}(\mathcal{B}_{1i}^{*},\mathcal{B}_{2i}^{*},\mathcal{B}_{3i}^{*})}\left(\begin{array}{ccc}
\mathcal{B}_{1i}^{*} & -\mathcal{B}_{2i}^{*} & \mathcal{B}_{3i}^{*}\end{array}\right).
\end{equation}

\end{document}